\documentclass[traditabstract]{aa}
\usepackage{graphicx}
\usepackage{txfonts}
\usepackage{natbib}
\usepackage{bm}

\bibpunct{(}{)}{;}{a}{}{,}

\begin{document}

\title{Compressive Sensing for Spectroscopy and Polarimetry}
\author{A. Asensio Ramos\inst{1,2} \& A. L\'opez Ariste\inst{3}}

\offprints{aasensio@iac.es}

\institute{
 Instituto de Astrof\'{\i}sica de Canarias, 38205, La Laguna, Tenerife, Spain; \email{aasensio@iac.es}
\and
Departamento de Astrof\'{\i}sica, Universidad de La Laguna, E-38205 La Laguna, Tenerife, Spain
\and
THEMIS, CNRS UPS 853, c/v\'{\i}a L\'actea s/n. 38200. La  Laguna,
Tenerife, Spain; \email{arturo@themis.iac.es}
}
\date{Received ; Accepted}
\titlerunning{}
\authorrunning{Asensio Ramos et al.}

\abstract{We demonstrate through numerical simulations with real data
the feasibility of using compressive sensing techniques for the acquisition of spectro-polarimetric
data. This allows us to combine the measurement and the compression process into one
consistent framework. Signals are recovered thanks to a sparse reconstruction
scheme from projections of the signal of interest onto appropriately chosen vectors, typically noise-like
vectors. The compressibility properties of spectral lines are analyzed in detail. The results shown
in this paper demonstrate that, thanks to the compressibility properties of spectral lines, it is 
feasible to reconstruct the signals using only a small fraction of the information that is measured
nowadays. We investigate in depth the quality of the reconstruction as a function of the amount of
data measured and the influence of noise. This change of paradigm also allows us to define new instrumental
strategies and to propose modifications to existing instruments in order to take advantage of compressive sensing techniques.}
\keywords{Techniques: spectroscopic --- Techniques: polarimetric --- Methods: observational --- Magnetic fields}

\maketitle


\section{Introduction}
\label{sec:intro}
Our present knowledge of the physical and magnetic properties of solar and stellar plasmas
owns a debt to the rapid development of spectro-polarimeters in the last decades. These
instruments use dispersive elements for the spectral analysis. Because visible/infrared
detectors are only sensitive to the intensity of light, modulators are used to encode the polarimetric information 
on intensity variations. Since most detectors are nowadays two-dimensional,
the recovery of two-dimensional spectro-polarimetric information is carried out using scanning
techniques: spatial in the case of long-slit spectro-polarimeters and spectral in the case of
Fabry-Perot-like spectro-polarimeters. According to the Nyquist-Shannon sampling theorem \citep{nyquist28,shannon49}, 
the correct sampling of a band-limited signal should be done at a rate equal to twice the bandwidth. In
other words, one should sample each resolution element (spectral and/or spatial) with two
pixels at least in order to be assure that all the frequencies in the bandwidth can
be observed. This theorem has been applied with faith during the last half century but one
should note that it is only valid for band-limited signals.

During the last few years, the emerging theory of compressive sensing 
\citep[CS;][]{candes06,donoho06} is showing that this sampling is indeed too restrictive when 
some details of the signal structure are known in advance. The interesting point of
the new CS paradigm is that, in many instances, natural signals have a structure
that is known in advance. For instance, stellar oscillations can be represented by sinusoidal
functions of different frequencies, images can be represented in a multiresolution analysis
using wavelets, etc. The key point is that, typically, only few elements of the
basis set in which we develop the signal are necessary for an accurate description
of the important physical information. The innovative character of CS is that this
\emph{compressibility} of the observed signals is inherently taken into account in the
measurement step, and not only in the post-analysis, thus leading to efficient measurement
protocols. Instead of measuring the full signal (wavelength variation of the Stokes
profiles in our case), under the CS framework, one measures a few linear projections of the signal
along some vectors known in advance and reconstructs the signal solving a non-linear problem.

A quick review of the literature shows us that signals arising in natural phenomena are typically
compressible (see JPEG\footnote{The name JPEG stands for Joint Photographic Experts Group, and is
a lossy compression format for images based on the application of sparsity-enhancing linear transformations.} 
compression and wavelet, curvelet or ridgelet decompositions of images, among
many others). Since this is also the case for astronomical data 
\citep[e.g.,][]{muhlmann_wavelet96,fligge_wavelet97,belmon02,polygiannakis03,dollet04,bernas04}, it has been suggested
that CS can be used to alleviate telemetry problems with space telescopes like
Herschel \citep{bobin_cs_herschel08} or Cassini \citep{belmon02} and to improve
existing techniques for the reconstruction of radio interferometric data \citep{wiaux_interferometry_cs09}. 
More specifically, several works also demonstrate 
that this is also true 
in the field of polarimetry \citep{socas_arturo_lites01,arturo_casini02,skumanich02,casini05}.
Following this idea, \cite{lites_jpeg02} have analyzed the effectiveness of using JPEG 
compression for reducing the amount of data that needs to be transferred through telemetry for
the Hinode space telescope. This option is currently in use in the mission \citep[see][]{tsuneta_hinode08}.
All these results and those of the analysis we carry out in this paper prompt
the interest of investigating the appropriateness of employing CS ideas to measure
spectro-polarimetric signals, specifically in space telescopes but also for ground-based
telescopes.
Through the use of these techniques, 
we anticipate an enhanced performance in terms of de-noising and data acquisition rates which eventually 
may have an impact on the choice of detector technologies and data transfer.

The outline of the paper is the following. Section \ref{sec:compressive_sensing} gives a brief description
of the CS paradigm, showing how signals can be recovered from a few linear projections and
the properties that such projections need to fulfill. An analysis of the compressibility of spectro-polarimetric
signals of interest is shown in \S\ref{sec:compressibility}. Section \ref{sec:recovery} presents
recovery examples, with an analysis of the influence of noise. Finally, \S\ref{sec:other} shows novel instrumental
strategies based on CS ideas, while the conclusions are presented in \S\ref{sec:conclusions}.

\section{Compressive sensing}
\label{sec:compressive_sensing}
\subsection{Theoretical considerations}
Because of the innovative character of CS, we give a brief description of the most important points, 
while more mathematical details are discussed in Appendix \ref{sec:appendix}. For a more 
in-depth description, we refer the reader to recent references
\citep[e.g.,][and references therein]{baraniuk07,candes08}.

The usage of compressive sensing techniques for the measurement of a signal $\mathbf{x}'$
represented as a vector of length $M$ is based on the following two key ideas:
\begin{itemize}
\item Instead of measuring the signal itself, one
measures the scalar product of the signal with carefully selected vectors:
\begin{equation}
\mathbf{y} = \mathbf{\Phi} \mathbf{x}' + \mathbf{e},
\label{eq:sensing}
\end{equation}
where $\mathbf{y}$ is the vector of measurements of dimension $N$, $\mathbf{\Phi}$ is 
an $N\times M$ sensing matrix and
$\mathbf{e}$ is a vector of dimension $N$ that characterizes the noise on the measurement process.
Note that the previous
equation describes the most general linear multiplexing scheme in which the number of
measurements $M$ and the length of the signal $N$ may differ. In the
standard multiplexing case, the number of scalar products measured equals the dimension
of the signal ($N=M$). Consequently, it is possible to recover the vector $\mathbf{x'}$ provided
that $\mathrm{rank}(\mathbf{\Phi})=N$, so that the problem is not
ill-conditioned. In other words, one has to verify that every row of the $\mathbf{\Phi}$ matrix 
is orthogonal with respect to every other row. 

\item The second key ingredient of CS is the assumption that the signal of interest is 
sparse in a certain basis set (or can be efficiently compressed in this basis set). Any compressible
signal\footnote{A signal is said to be compressible (or quasi-sparse) if 
it is possible to find a basis for which the projection coefficients along the vectors
of the basis reordered in
decreasing magnitude decay in absolute value like a power-law.} can be written, in general, in the following way:
\begin{equation}
\mathbf{x}' = \mathbf{W}^T \mathbf{x},
\label{eq:basis_decomposition}
\end{equation}
where $\mathbf{x}$ is a $K$-sparse\footnote{A vector is $K$-sparse if only $K$ elements of
the vector are different from zero.} vector of size $M$ and $\mathbf{W}^T$ is the
transpose of a $M \times M$ transformation matrix associated with the basis set in which 
the signal is sparse. For instance, $\mathbf{W}$ can be the Fourier matrix if the signal
$\mathbf{x'}$ is the combination of a few sinusoidal components. Other transformations of
interest are the wavelet matrices or even empirical transformation matrices like those
found using principal component analysis. 
\end{itemize}

The combination of the those ingredients leads to the multiplexing scheme:
\begin{equation}
\mathbf{y} = \mathbf{\Phi} \mathbf{W}^T \mathbf{x} + \mathbf{e},
\label{eq:sensing2}
\end{equation}
with the hypothesis that $\mathbf{x}$ is sparse, which renders
CS feasible. It has been demonstrated by \cite{candes06} that, even if 
$\mathrm{rank}(\mathbf{\Phi} \mathbf{W}^T) \ll N$ (we have much fewer equations than
unknowns), the signal $\mathbf{x}$ can be recovered with overwhelming probability when using
appropriately chosen sensing matrices $\mathbf{\Phi}$.

\begin{figure}
\includegraphics[width=\columnwidth]{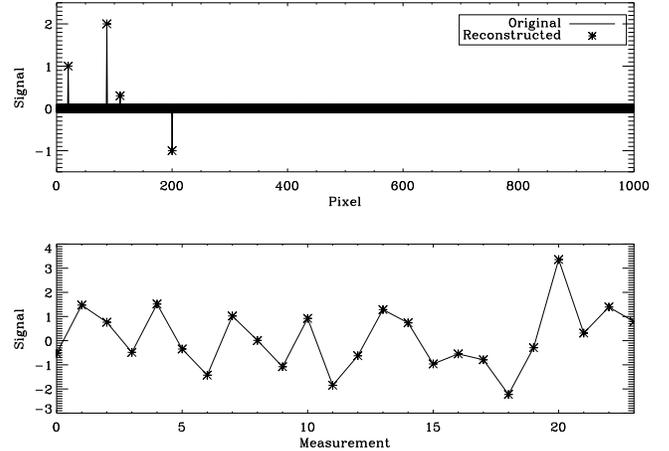}
\caption{Test showing how compressive sensing works for detecting a sparse signal. The upper
panel shows a sparse signal made of four spikes of very short duration. The lower panel presents
24 measurements built as the scalar product of the signal with Gaussian random noise. The 
stars show the reconstructed signal, showing that perfect recovery is possible with only
a very small fraction of the measurements-}
\label{fig:test_simple}
\end{figure}

When the number of equations is less than the number of unknowns, 
it is usual to solve Eq. (\ref{eq:sensing2}) using least-squares methods that try to minimize the 
$\ell_2$ norm\footnote{The $\ell_n$ norm of a vector is
given by $\parallel \mathbf{x} \parallel_n = (\sum_i |x_i|^n)^{1/n}$ if $n > 0$. The $\ell_0$ pseudo-norm 
is given by the number of non-zero elements of $\mathbf{x}$.} of the residual. This is
usually accomplished using techniques based on the singular value decomposition 
\citep[see, e.g.,][]{numerical_recipes86}. However, such minimization is known
to return non-sparse results \citep[e.g.,][and references therein]{romberg08}. 
A more appropriate solution to Eq. (\ref{eq:sensing2})
is to look for the vector with the smallest $\ell_0$ pseudo-norm (the number of non-zero elements 
of the vector) that fulfills the equation \citep{candes06}:
\begin{equation}
\min_{\mathbf{x}} \parallel \mathbf{x} \parallel_0 \textrm{subject to} 
\parallel \mathbf{y} - \mathbf{\Phi} \mathbf{W}^T \mathbf{x} \parallel_2 < \epsilon,
\label{eq:l0_minimization}
\end{equation}
where $\epsilon$ is an appropriately small quantity.
The solution of the previous problem is, in general, not computationally feasible. 
However, \citep{candes06,candes_2_06} demonstrated
that, if the matrix $\mathbf{\Phi} \mathbf{W}^T$ fulfills certain conditions described in Appendix \ref{sec:appendix}
\citep{candes_2_06}, the problem
\begin{equation}
\min_{\mathbf{x}} \parallel \mathbf{x} \parallel_1 \textrm{subject to} 
\parallel \mathbf{y} - \mathbf{\Phi} \mathbf{W}^T \mathbf{x} \parallel_2 < \epsilon,
\label{eq:l1_minimization}
\end{equation}
is equivalent to that of Eq. (\ref{eq:l0_minimization}). The advantage is that very
efficient numerical methods exist for the solution of such problem (the one used in this
paper is described in Appendix \ref{sec:appendix}).

Figure \ref{fig:test_simple} shows a very simple example that summarizes the essence of compressive
sensing. The upper panel presents a signal (solid line) that is 4-sparse in the basis of Dirac delta
functions. Instead of measuring the full length of the signal, the lower panel of the figure shows 
a very small amount of scalar products of the signal with a Gaussian random matrix. The stars present
the reconstructed signal using only 24 such measurements. Since this is a noiseless example, perfect
reconstruction is obtained.

\begin{figure*}
\includegraphics[width=0.49\textwidth]{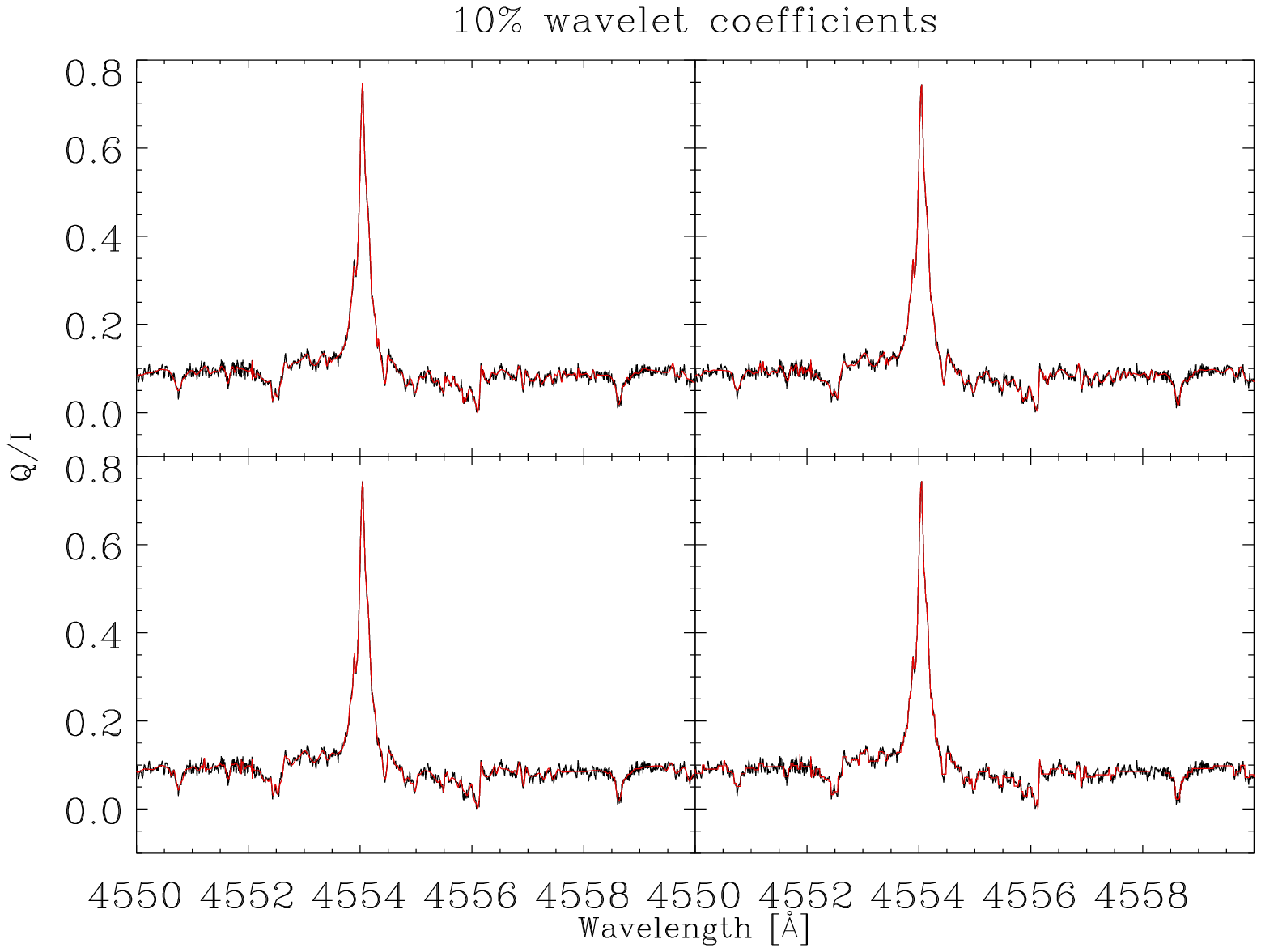}
\includegraphics[width=0.49\textwidth]{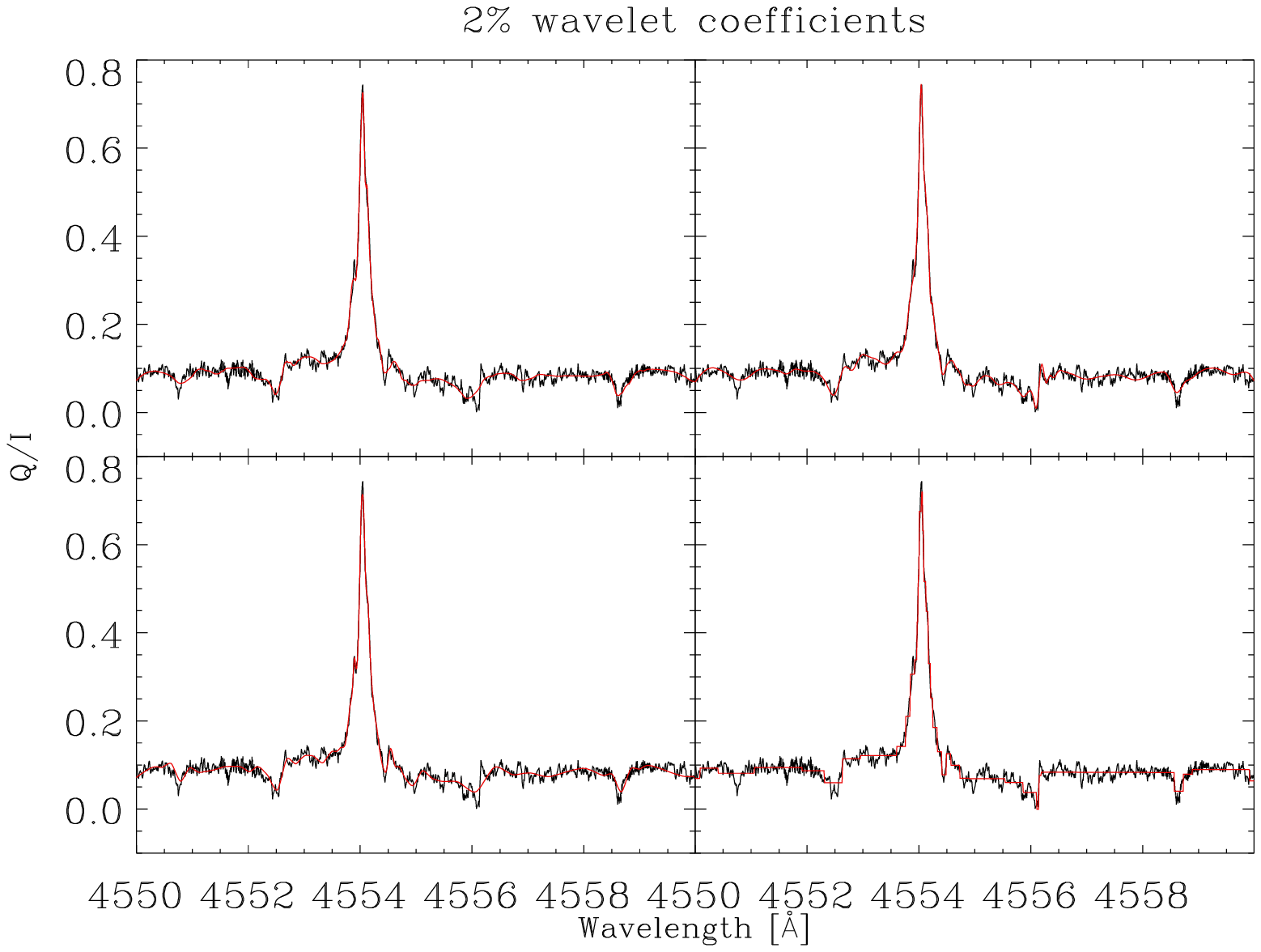}
\includegraphics[width=0.49\textwidth]{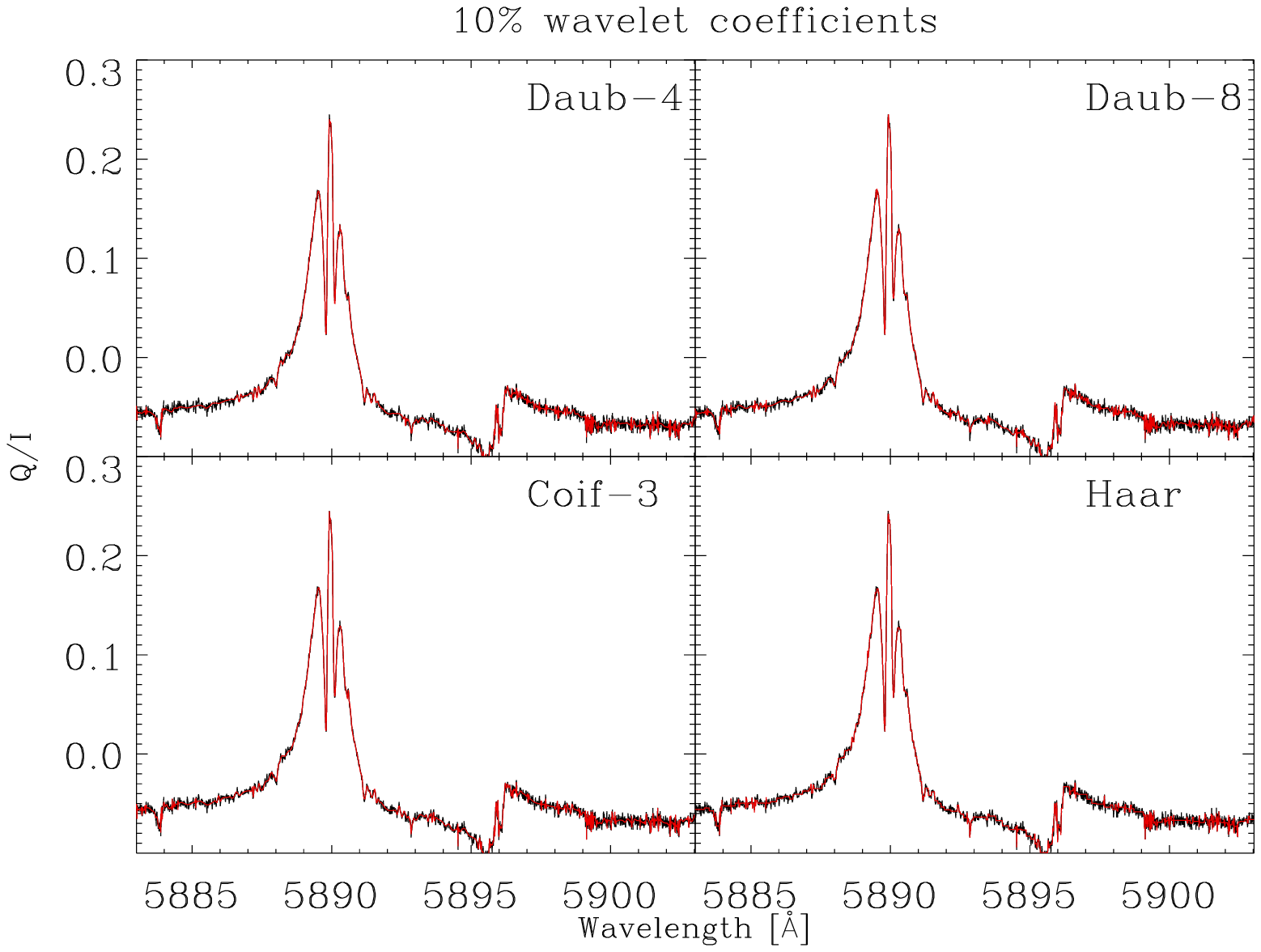}
\includegraphics[width=0.49\textwidth]{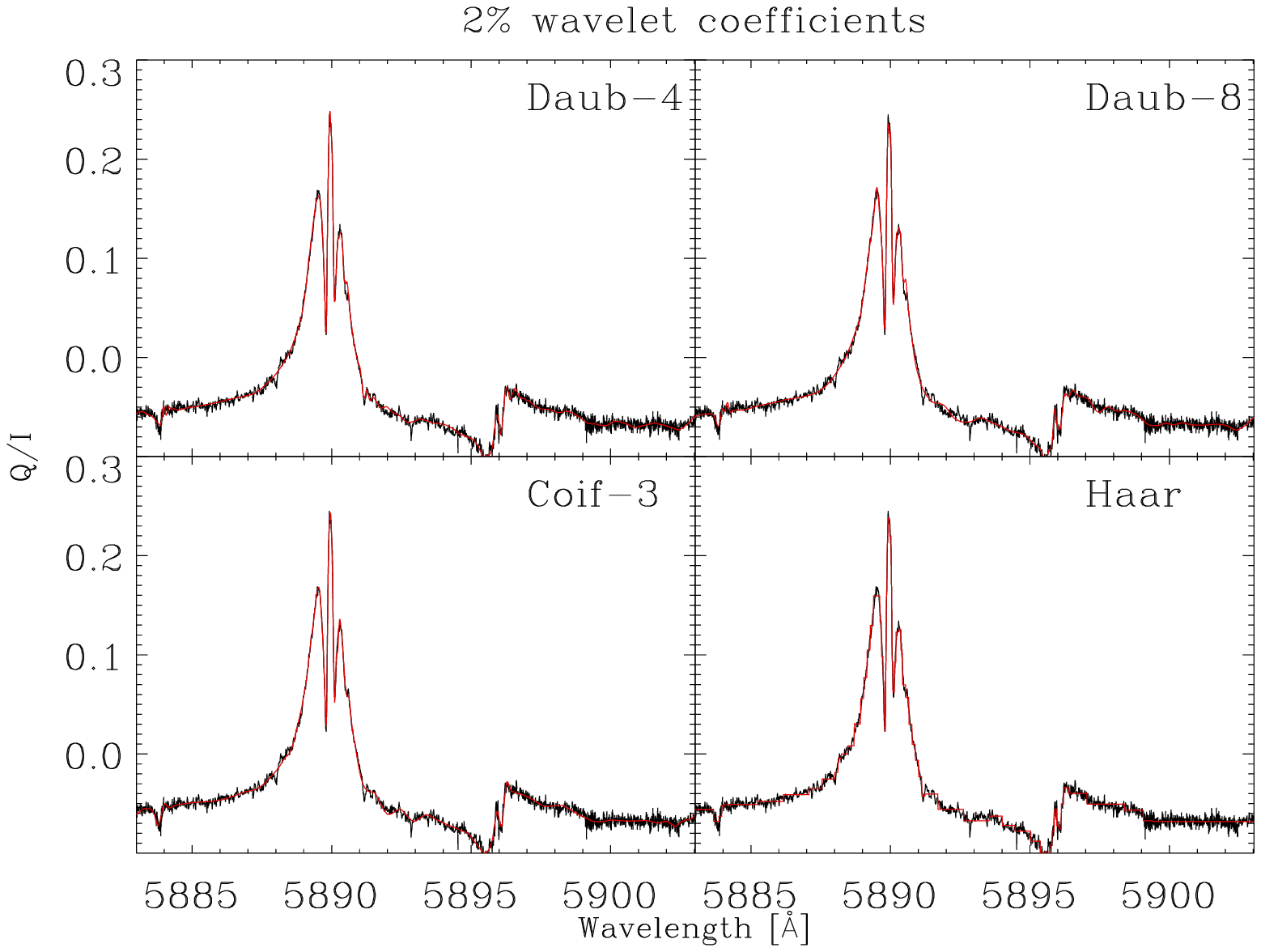}
\includegraphics[width=0.49\textwidth]{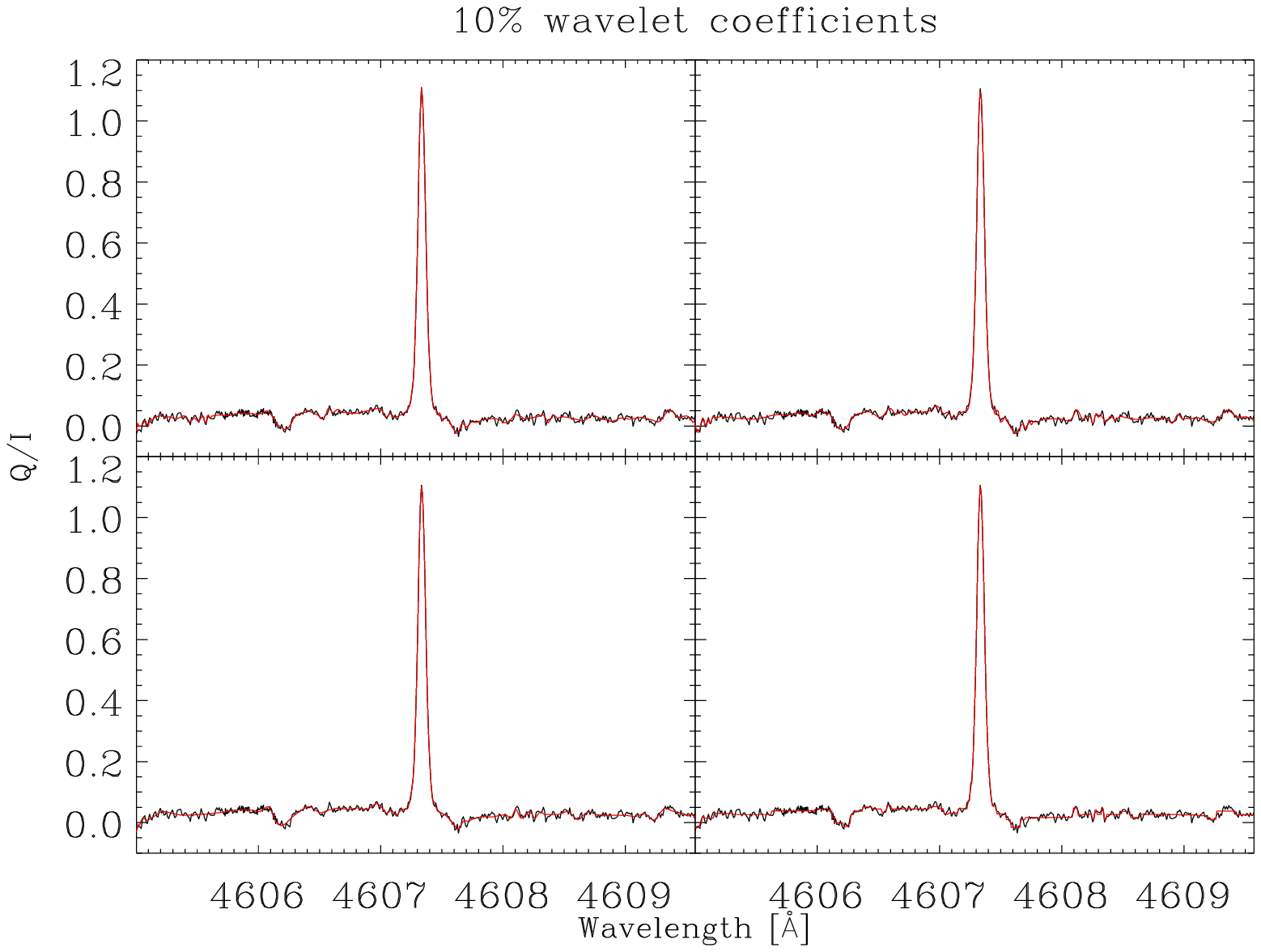}
\includegraphics[width=0.49\textwidth]{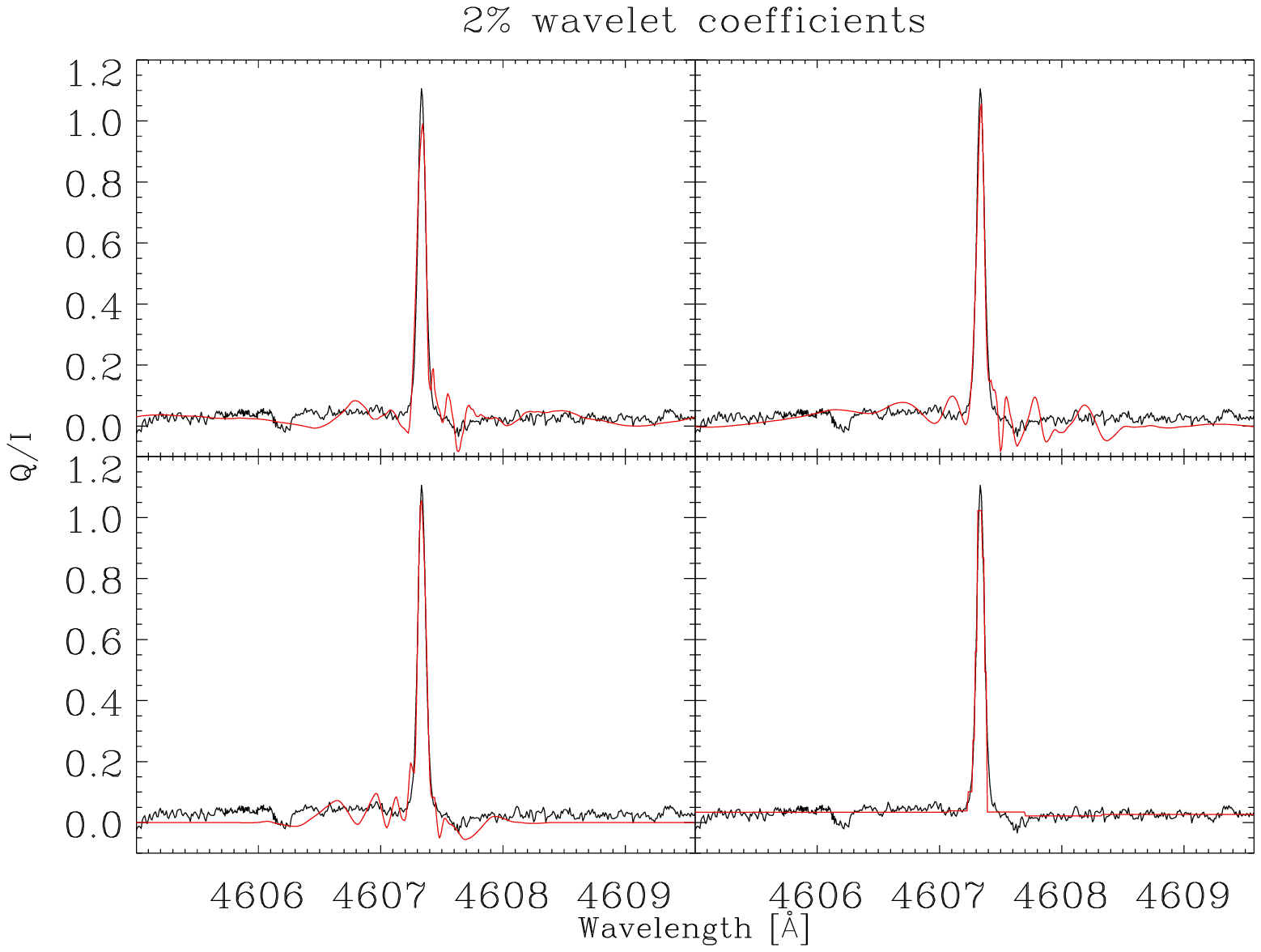}
\caption{Test showing the reconstruction of three widely known domains of the second solar spectrum: the top panels show
a region around the Ba \textsc{ii} D$_2$, the middle panels a region around the Na \textsc{i} doublet at 5890 \AA\
while the lower panel presents the region around the Sr \textsc{i} line at 4607 \AA.
Reconstruction using different wavelets and different thresholds are displayed. In each panel,
the left panel presents how the reconstructed spectrum (red line) compares with the
original spectrum using Daubechies-4, Daubechies-8, Coiflet-3 and Haar wavelets when 90\% of the wavelet coefficients
are set to zero. The right panels show the results when only 2\% of the wavelet coefficients are retained. These
results demonstrate that the structure of the lines is nicely recovered with such a few number of wavelet coefficients.
Only in the case of the Sr \textsc{i}, we find spurious ripples (except in the Haar case) due to the 
loss of information.}
\label{fig:reconstruction_full}
\end{figure*}

\begin{figure*}
\includegraphics[width=0.49\textwidth]{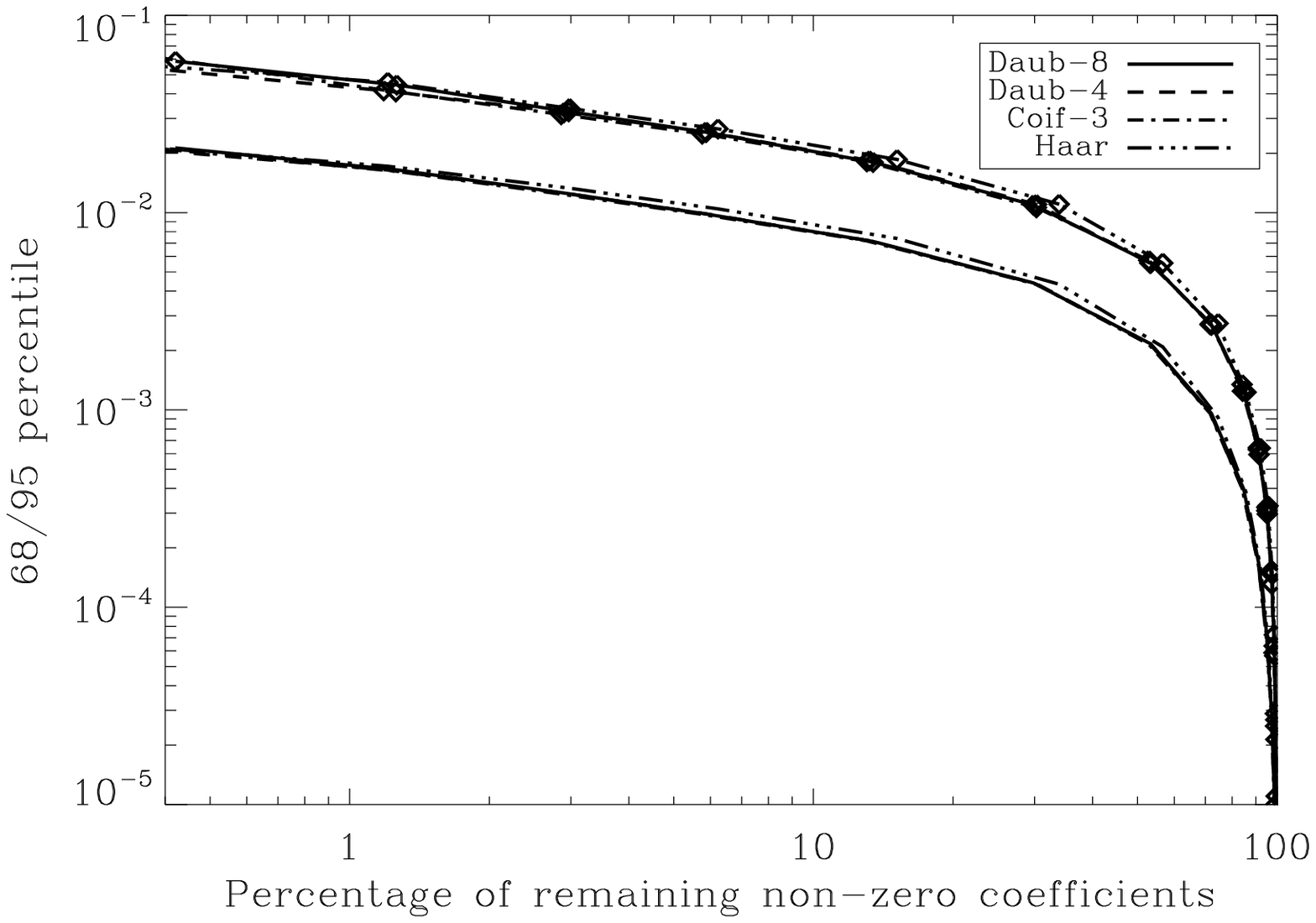}
\includegraphics[width=0.49\textwidth]{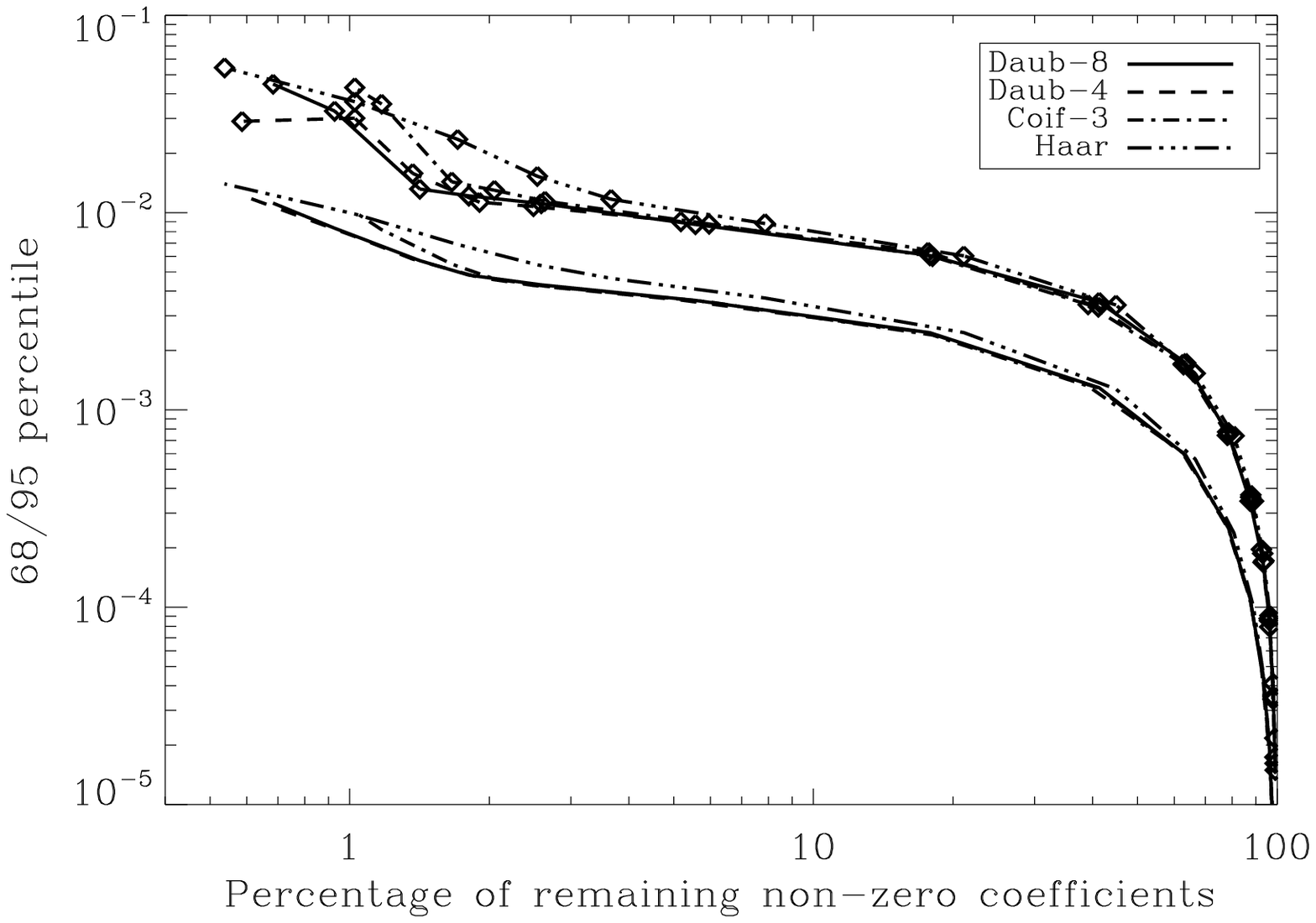}
\caption{Percentiles 68\% (lines) and 95\% (lines+symbols) of the differences between the true signal and the reconstructed
signal when retaining a different percentage of the wavelet coefficients. The left panel corresponds
to the data close to the Ba \textsc{ii} D$_2$ line, while the right panel is focused on the reconstruction
of the Na \textsc{i} doublet.}
\label{fig:percentile}
\end{figure*}

\subsection{Practical considerations}
The CS framework offers several advantages over standard measurement paradigms. On the one hand,
since one only measures linear combinations of the signals, the flux of information
that any sensor has to deal with is usually much smaller (thanks to compression). This is probably
of secondary interest for ground-based instruments since the infrastructure to cope with such large
fluxes of data is available. However, this could be of interest for space-borne instruments,
where the flux of data is limited by telemetry. As a sub-product of
the simplification on the measurement, the reconstruction of the signal is much more time consuming,
but can be done efficiently a-posteriori without affecting the measurement process. On the other hand, if 
appropriate sensing matrices are chosen, the measurement process can be considered universal and does not depend
on the exact basis set in which the data is sparse. In other words, one first measures 
projections and this assures that the data can be reconstructed a-posteriori
if the data is sparse in any (unknown a-priori) basis set.

Ground-based instruments may draw advantages from the increased cadence at which data is acquired. An 
instrument measuring spatial and spectral information with a 2D detector is forced to scan one of 
the extra dimensions of the data space, either the spectra in filter instruments or one 
of the spatial dimensions in spectrograph instruments. The CS framework may diminish the number of 
measurements in the spectrum space thus shortening the time in which the 3D data cube is acquired by 
the instruments. Boosting data acquisition rates is always advantageous from both the 
point of view of a time evolving observational target (as solar structures) or from the point of view 
of deformations or aberrations introduced by atmospheric turbulence (seeing).

\begin{figure*}
\includegraphics[width=0.49\textwidth]{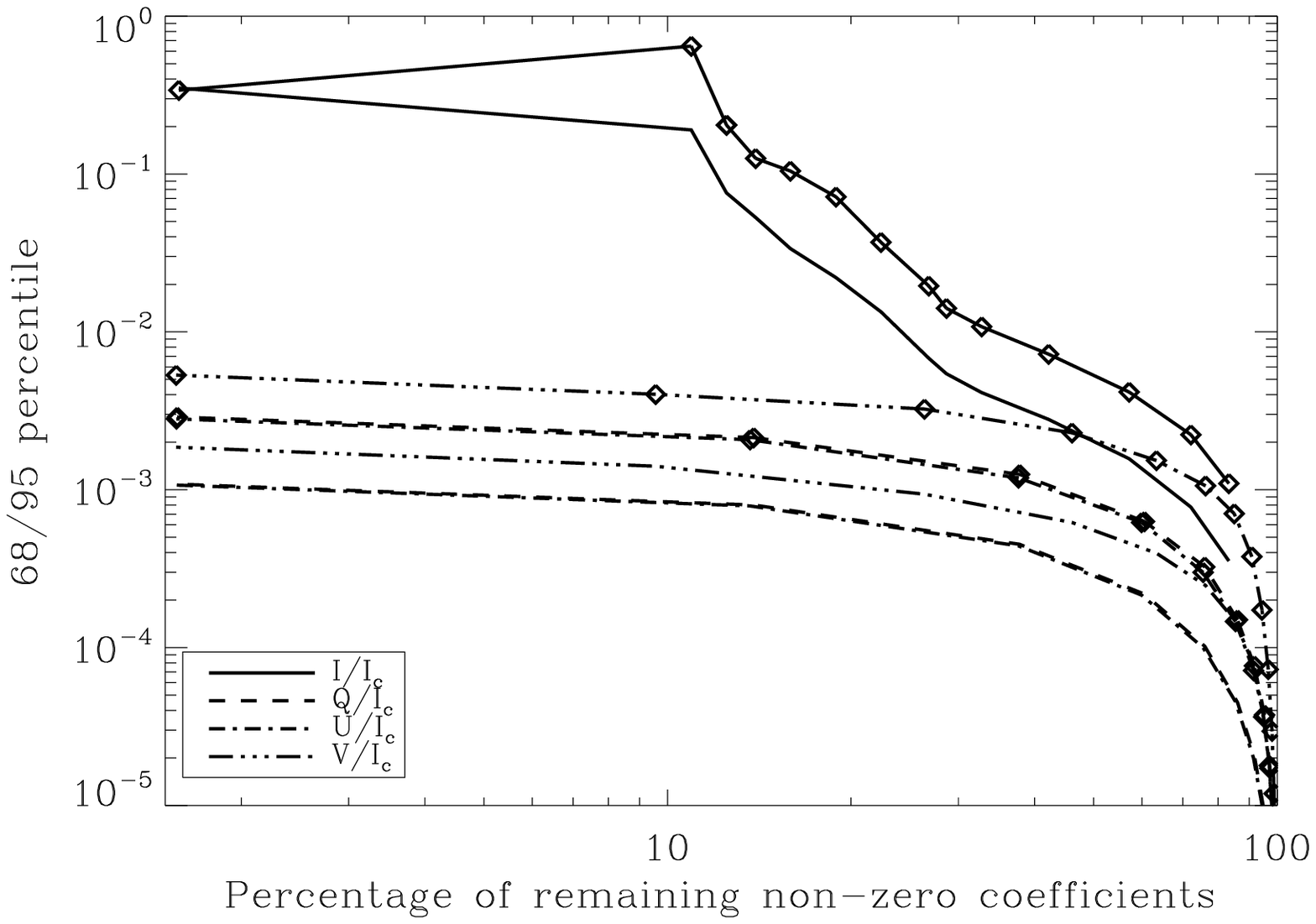}%
\includegraphics[width=0.49\textwidth]{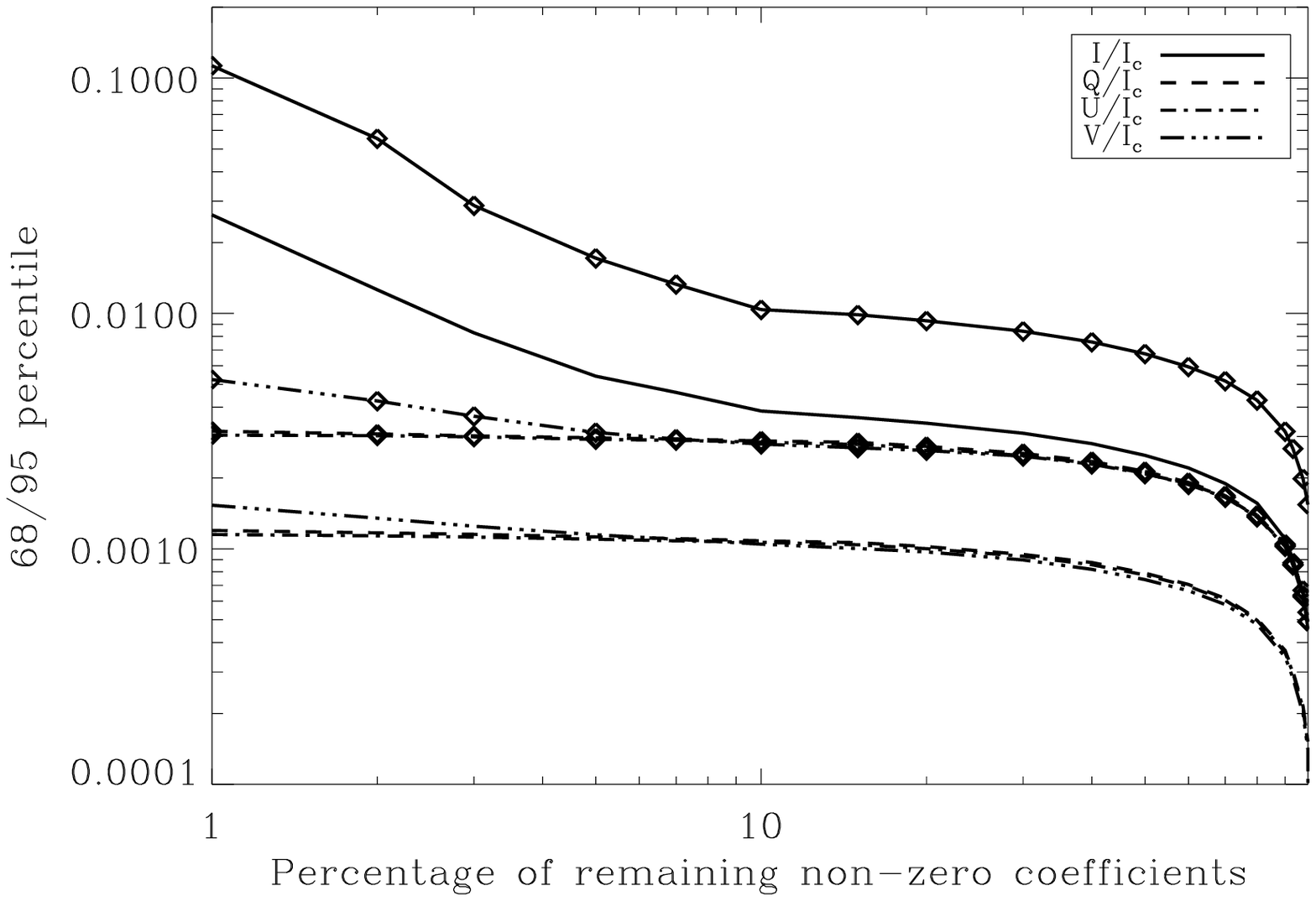}
\caption{Percentile 68\% (lines) and 95\% (lines+symbols) of the differences between the true signal and the reconstructed
signal for a Hinode observation. The left panel shows the case using a Daubechies-8 wavelet, while the
right panel presents the results using the PCA empirical basis set. The noise level in Stokes $Q$, $U$ and $V$ is close to 1.6$\times$10$^{-3}$
in units of the continuum intensity. Therefore, less than 5\% of the PCA eigenvectors are needed in order
to reconstruct the profiles to this noise level.}
\label{fig:percentile_zeeman}
\end{figure*}

\section{Compressibility of the signals}
\label{sec:compressibility}
As reported in \S\ref{sec:compressive_sensing}, any signal amenable to compressive sensing has to be
sparse or compressible in some basis set.
The purpose of this section is to test to what extent
polarimetric signals are compressible \citep{asensio_dimension07}. We focus mainly on
linear and circular polarization profiles, although our results can be extended to
standard spectroscopic observations without effort. We present results for signals produced by scattering processes
and for signals produced by the Zeeman effect under the presence of a magnetic field. Concerning
the basis set in which the signals are sparse, we focus on the wavelet family for the case of 
scattering polarization and on the principal component analysis \citep[PCA;][]{loeve55} decomposition 
for the Zeeman case. These are just two 
possible candidates and we want to stress that it is advantageous to analyze each
case in detail in order to find the basis set in which the signal is as sparse as possible.

\subsection{Sparsity of the Second Solar Spectrum}
The wavelength variation of the fractional linear polarization $Q/I$ measured very close to the
solar limb is nowadays usually known as the second solar spectrum.
Its name comes from the large wavelength variability, in some sense
comparable to the standard Fraunhoffer intensity spectrum \citep[see][]{gandorfer_atlas1_00,gandorfer_atlas2_02,gandorfer_atlas3_05}.
Examples of the variability are shown in Fig. \ref{fig:reconstruction_full}, where many of the peaks detected correspond 
to specific spectral lines. Certain lines produce very conspicuous signals like the neutral sodium doublet shown in the
middle panel of Fig. \ref{fig:reconstruction_full}.

It is apparent that the spectrum of $Q/I$ cannot be, in general, considered to
be sparse in the wavelength domain because it is composed of broad peaks with a large wavelength variability.
However, driven by the typical shape of the line profiles, we analyze the sparsity
properties of the second solar spectrum when decomposed on the wavelet domain. To this end, we select
standard wavelet mothers that are widely used in other application \citep[e.g.,][]{ripples01}: the Daubechies family, the
Coiflet family and the Haar family. The discontinuous Haar family is appropriate for decomposing
pixel-based data. The other families produce smoother approximation to the data. It is left for the
future to analyze the potential of other families, especially non-orthogonal 
redundant wavelets \citep[e.g.,][]{starck97,fligge_wavelet97} or Hermite functions \citep{hermite_deltoro03}.

We carry out an experiment for characterizing the compressibility of the second solar spectrum. 
We select a large piece of the spectrum in which many small signals are present, together
with a strong signal produced by the D$_2$ line of Ba \textsc{ii}. The spectrum is shown in 
black solid lines in Fig. \ref{fig:reconstruction_full}. The full spectrum
is wavelet-decomposed (the wavelet of choice is shown in each panel) and thresholded so that only a
certain number of wavelet coefficients survive, while the rest of coefficients are set to zero. This is
an efficient way of compressing the signal provided that the thresholding fundamentally cancels noise
and leaves the signal unperturbed. The upper left panel of Fig. \ref{fig:reconstruction_full} shows what
happens when only 10\% of the wavelet coefficients are maintained, while the right panel indicates
the behavior after setting to zero 98\% of the coefficients. Since the zero coefficients are not 
necessary in the reconstruction, this thresholding leads to an important compression of the signal.
The signal is then reconstructed using the inverse wavelet transform.
We note that even in the case of only 2\% of the coefficients, the important signals are nicely recovered
while the noisy part of the spectrum is largely reduced. Apparent from the figure is the fact that 
the behavior is very similar for all the wavelet families we have tested, although the computing
times are different, being larger for wavelets with a larger number of non-vanishing moments \citep[e.g.,][]{ripples01}.
This can be an issue that should be taken into account depending on the 
balance between the desired smoothness of the reconstruction and the computing time.

Other examples are shown in the middle and lower panels of 
Fig. \ref{fig:reconstruction_full}, for the case of the Na \textsc{i} doublet at 5890 \AA and
the Sr \textsc{i} line at 4607 \AA, respectively. The first one presents a case in which 
low-frequency (the large scale quantum interference between the two lines
of the doublet) and high-frequency information (the large variability of the profiles close to the core
of the line) coexist. This poses an interesting problem to any compression method because it has to
retain low- and high-frequencies simultaneously. Apparently, the wavelet compression does a good job
on this multiplet and all important details can be retrieved even with only 2\% of the coefficients.
The lower panels of Fig. \ref{fig:reconstruction_full} show the case of the Sr \textsc{i} line at 
4607 \AA, which is a very strong signal embedded in a quasi-flat continuum. In this case, reconstructing
only with 2\% of the coefficients gives a bad representation of the true underlying signal. Some
ripples appear when using Daubechies and Coiflet wavelets on the quasi-continuum, although the
amplitude of the signal is still correctly recovered. The reconstruction with the Haar wavelet gives
a very good representation of the Sr \textsc{i} line but the depolarizations in the red wing of
the line and at 4606.3 \AA\ are not correctly recovered. The reconstruction with 10\% of the coefficients
is almost perfect.

We have measured the quality of the reconstruction
using the 68\% and 95 \% percentiles of the difference between the original and the reconstructed
signal. The value of these quantities versus the percentage of remaining non-zero coefficients is shown in
Fig. \ref{fig:percentile}. The left panel has been obtained with the data from the Ba \textsc{ii} D$_2$
line, while the right panel is associated with the Na \textsc{i} data.
We have verified that the distribution of differences is close to normal except in
the cases in which the reconstruction is done with too few coefficients. Therefore, the 68\% and 95\% percentiles
are close to the standard deviation and twice the standard deviation, respectively. The lines without symbols 
present the 68\% percentile, showing that differences are in both cases below 10$^{-2}$ when retaining
just 10\% of the profiles. For $Q/I$ signals that are on average at the level of $\sim 0.1$, we find that relative
errors are typically below 10$^{-3}$. The 95\% percentile gives relative errors slightly above 10$^{-3}$ for
these cases. The results tend to indicate that differences among wavelet families are relatively small.

\subsection{Sparsity of Zeeman signals}
\label{sec:sparsity_zeeman}
Under the presence of a sufficiently strong magnetic field, the Zeeman effect usually controls the emergent
observed polarization. In order to test for the compressibility properties in the Zeeman-dominated
case, we use a dataset obtained with the Solar Optical Telescope/Spectro-Polarimeter SOT/SP \citep{lites_hinode01} 
aboard \emph{Hinode} \citep{kosugi_hinode07}. The properties of this dataset are
typical of what one should encounter in the future if the CS techniques that we propose
here are applied to future space missions. In this case, we test for two different basis
set for compressibility in Fig. \ref{fig:percentile_zeeman}. The first is the universal Daubechies-8 
wavelet (left panel). The results we show are not very sensitive to the specific 
chosen wavelet and are representative of the general behavior. The second is the empirical basis 
set obtained using the PCA decomposition (right panel).

\begin{figure*}
\includegraphics[width=0.49\textwidth]{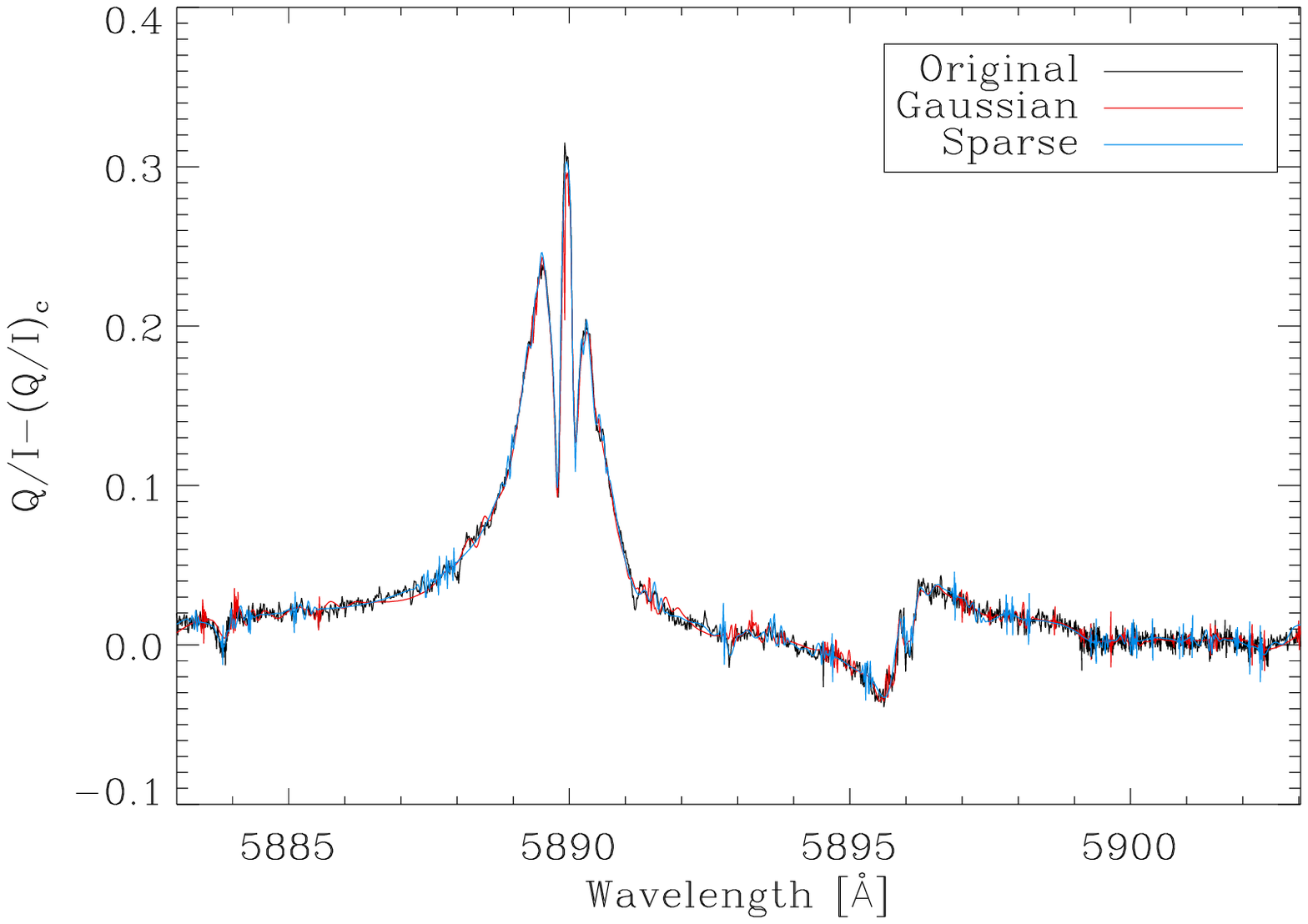}
\includegraphics[width=0.49\textwidth]{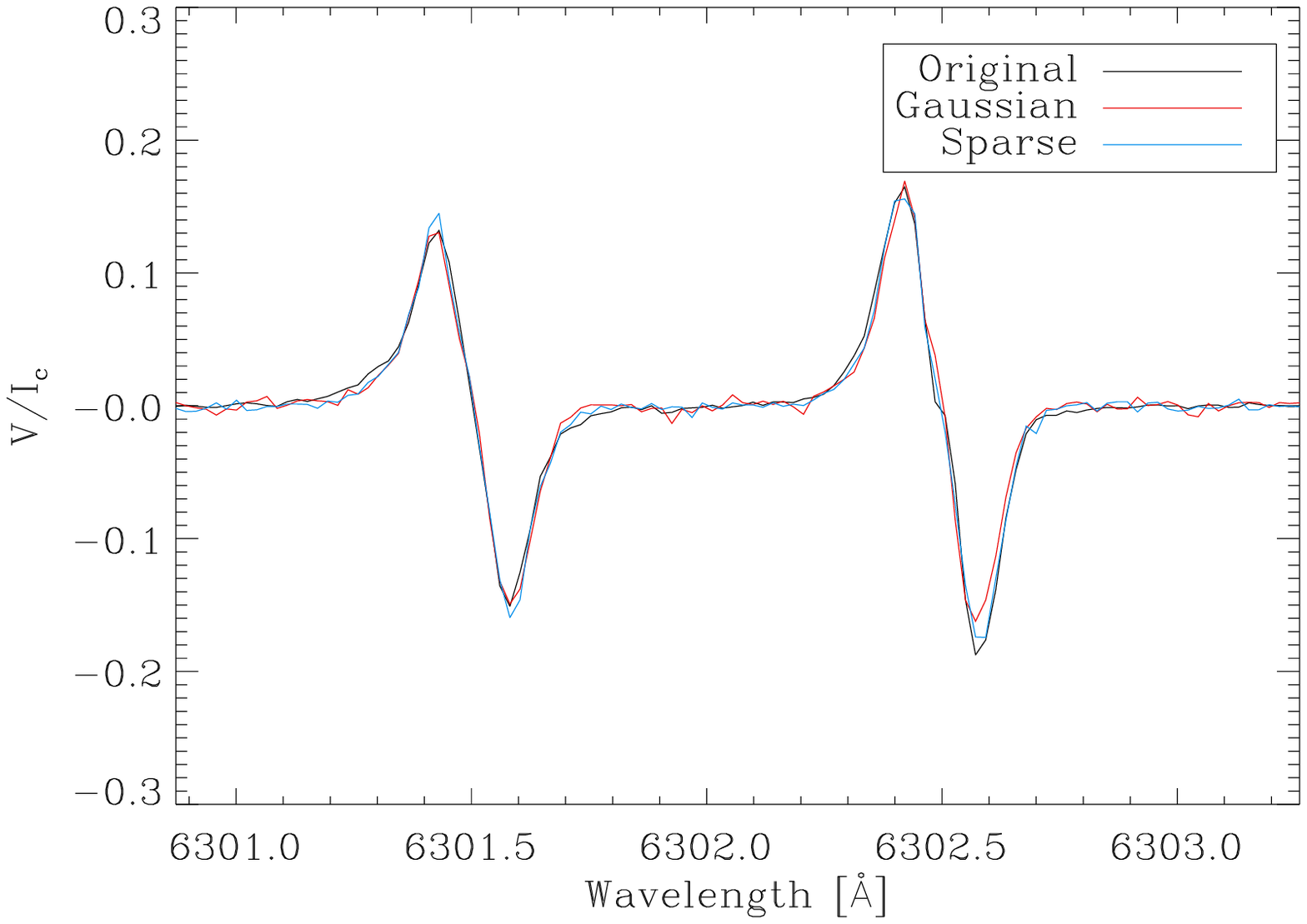}
\includegraphics[width=0.49\textwidth]{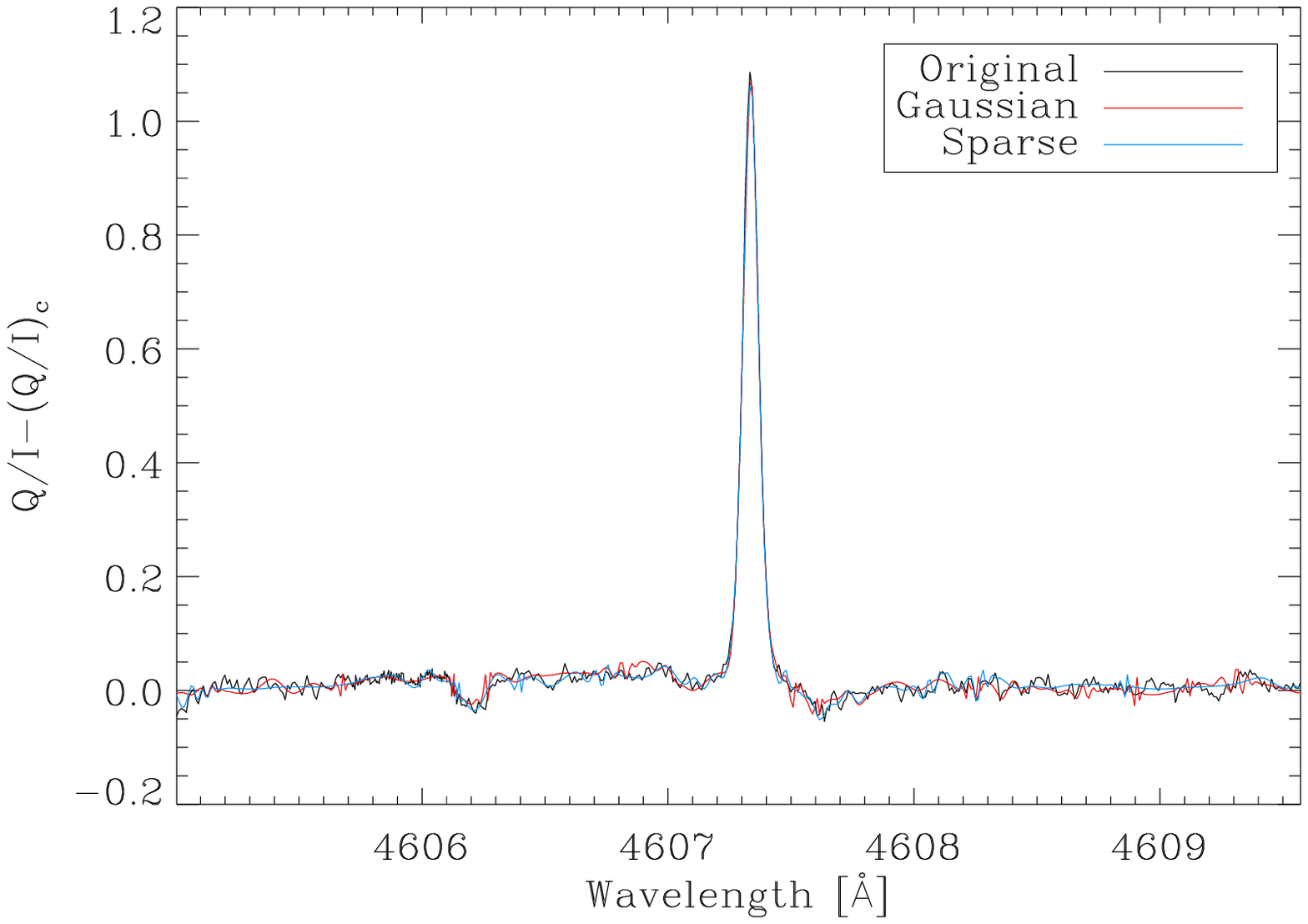}
\includegraphics[width=0.49\textwidth]{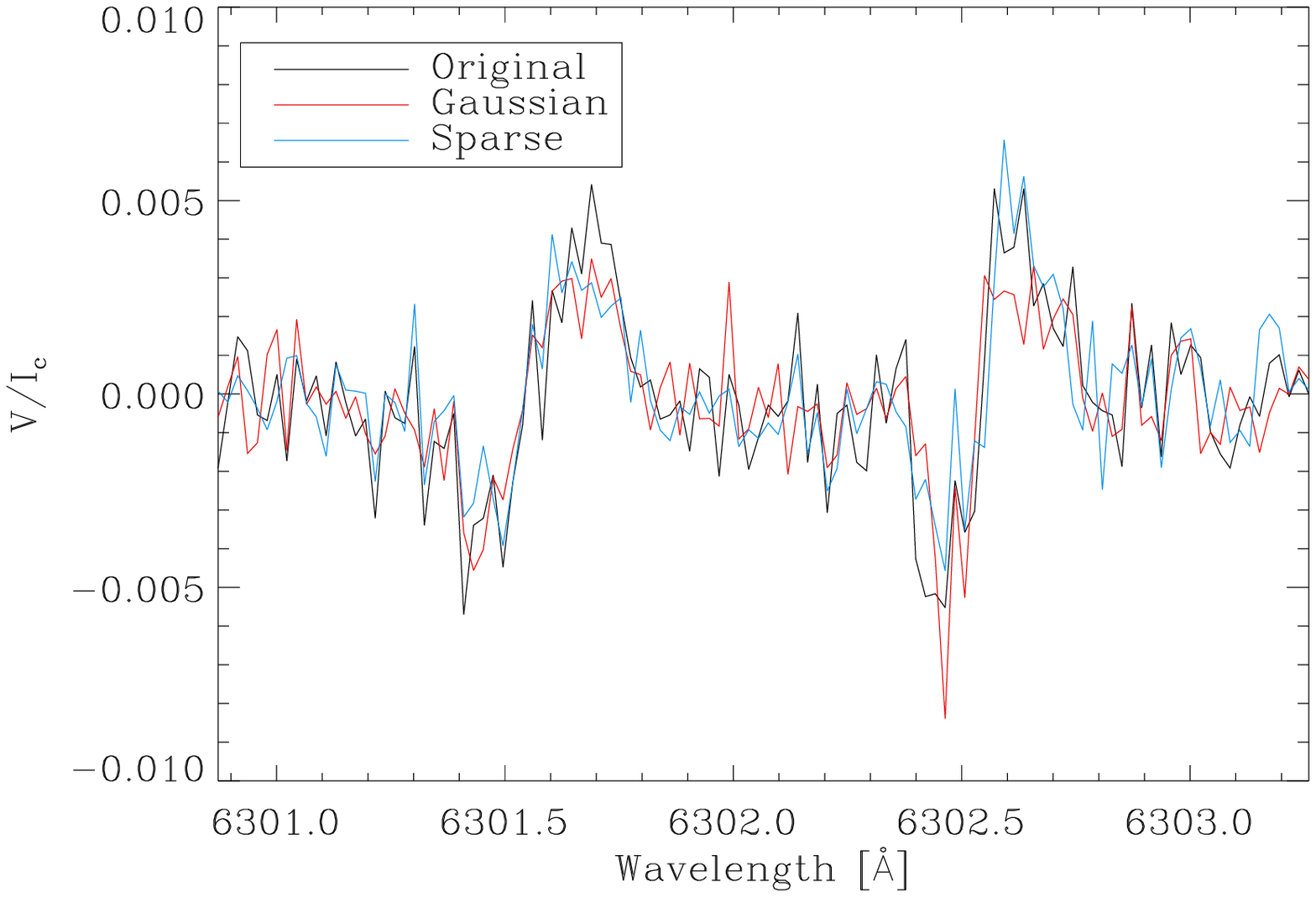}
\caption{Examples of CS reconstructions of four different polarization signals. The left panels
present the results for signals produced by scattering polarization when observed close to the limb.
Reconstruction is done using Daubechies-8 wavelets. The right panels show reconstructions of 
Zeeman signals observed with Hinode. Reconstruction is done using a universal PCA basis. Additionally, 
we show the difference on the reconstruction using two different sensing matrices: a Gaussian matrix 
with zero mean and inverse variance
equal to the sparsity of the signal and a sparse binary sensing matrix with 10 non-zero elements per measurement.
In general, we find a better behavior for the sparse binary matrix than for the random full matrix.}
\label{fig:reconstruction_Na_Sr_Zeeman}
\end{figure*}

The results, that present the 68\% and 95\% percentiles of the distribution of the difference between the
exact and the reconstructed profiles, indicate clearly that signals are again compressible. The best
results are, obviously, obtained with the PCA basis set, because the eigenvectors are empirically
constructed to maximize the sparsity of the signal (only a few eigenvectors are necessary to reconstruct
the signal without noise). 

The main problem with the PCA basis set is that it is obtained empirically. Consequently, strictly
speaking, one is not able to use the PCA basis set in a CS framework because a-priori the basis
set is not known. However, according
to \cite{skumanich02}, some universality properties of the PCA eigenvectors can be
demonstrated when many profiles are included in a database. For this reason, we have tested that
profiles observed with Hinode can be nicely compressed with the eigenvectors recovered from
a completely different dataset. The reason is that the same physical effects are controling the
signals in both datasets. This opens the possibility of using some kind of universal
PCA basis for compressing Zeeman-dominated data. This basis set will surely contain details of 
the spectral lines that are present in the majority of the observed profiles and that can be
hardly recovered with fixed basis sets like wavelets.

\section{Signal recovery}
\label{sec:recovery}

\subsection{Examples}
\begin{figure}[!b]
\includegraphics[width=\columnwidth]{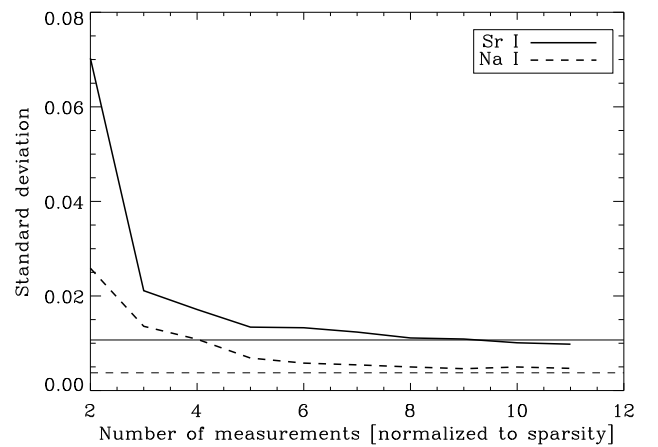}
\caption{Standard deviation of the difference between the recovery and the original signal
for different number of measurements. The reconstruction is done using the sparse binary matrix
and the Daubechies-8 wavelet.}
\label{fig:recovery_vs_meas}
\end{figure}

Since we have demonstrated that the polarimetric signals are compressible, the CS framework
can be used to measure such signals. We give here a few examples using different sensing matrices. The first ones are shown
in the left panel of Fig. \ref{fig:reconstruction_Na_Sr_Zeeman}. We present the reconstruction of the Na \textsc{i}
and Sr \textsc{i} signals analyzed in \S\ref{sec:compressibility} with two different sensing
matrices: (i) a Gaussian matrix with elements extracted from the $N(0,1/K)$ distribution \citep{candes06}, with
$K$ being the sparsity of the signal and (ii) a binary sparse sensing matrix with only 10
non-zero elements per measurement \citep{berinde_indyk08}. Obviously, the binary sparse matrix has two advantages over
the Gaussian matrix. First, the number of non-zero elements is very small as compared to the
size of the matrix and efficient sparse storage and computational methods can 
be used \citep[e.g.,][]{numerical_recipes86}. In our case, the sparse 
matrix contains less than 2\% of the elements different from zero. Second, the binary matrix
is easier to implement on hardware using, for instance, micro-mirrors. Only 5\% of the elements
of the solution vector are allowed to be non-zero for the case of Na \textsc{i} and 10\% for
the case of Sr \textsc{i}, according to the results presented in Fig. \ref{fig:reconstruction_full}. 
The number of measurements used is $6K$ for the Sr \textsc{i} line
and $8K$ for the Na \textsc{i} doublet, roughly in accordance with Eq. (\ref{eq:nmeas}), while the 
reconstruction is done using the Daubechies-8 wavelet. The results show that a good recovery is
possible in the two cases, with the advantage that, since sparsity is inherent to the reconstruction,
noise is largely reduced in the reconstruction. In order to show how the technique behaves with
the number of measurements, we show in Fig. \ref{fig:recovery_vs_meas} the standard deviation of the
difference between the reconstructed and original signal versus the number of measurements (normalized
to the sparsity of the vector). The horizontal lines indicate an estimation of the noise level in
the observations obtained as the standard deviation of a portion of the continuum. Note that,
when the number of measurements is not large enough, the reconstruction does not work. However, as soon
as condition (\ref{eq:nmeas}) is fulfilled, reconstruction works properly.

Other examples are shown on the right panel of Fig. \ref{fig:reconstruction_Na_Sr_Zeeman}. Stokes $V$
profiles picked at two positions of an observation carried out with Hinode on February 27, 2007.
The Fe \textsc{i} doublet at 630 nm with amplitudes typical of active regions (upper panel)
and quiet Sun (lower panel) are shown in black lines. The reconstructed signals using the same
sensing matrices as above are shown in red and blue. The universal PCA basis is used and only
11 of such eigenvectors are used, roughly 10\% of the full basis set. The number of measurements
is $6K$, in accordance with Eq. (\ref{eq:nmeas}). A good recovery is possible even for profiles
whose amplitude is close to the noise level. The prior information encoded in the sparse reconstruction
produces that noise is slightly reduced with respect to the original profile. This is similar to what
one would find after carrying out a PCA filtering of the data, but the filtering is encoded
inside the measurement technique \citep[e.g.,][]{marian08}. The fundamental reason for this is that,
while true signals produce sparse signatures, noise destroys the sparsity to some degree. Since the reconstruction
is done enhancing sparsity, it is not possible to recover noise and a filtering is carried out
as a side effect of the reconstruction.

\subsection{Sensitivity to noise}
\begin{figure*}
\includegraphics[width=0.49\textwidth]{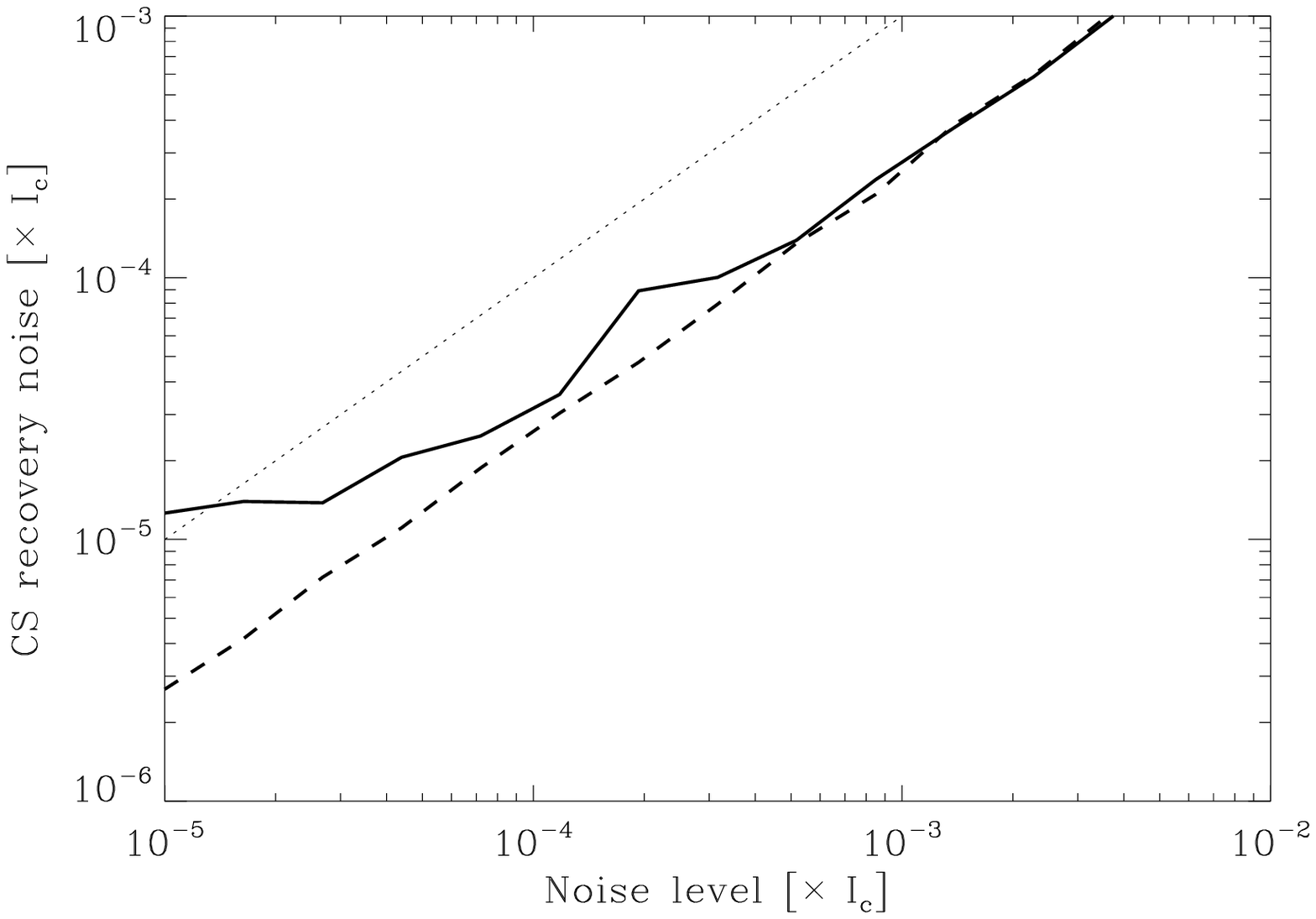}
\includegraphics[width=0.49\textwidth]{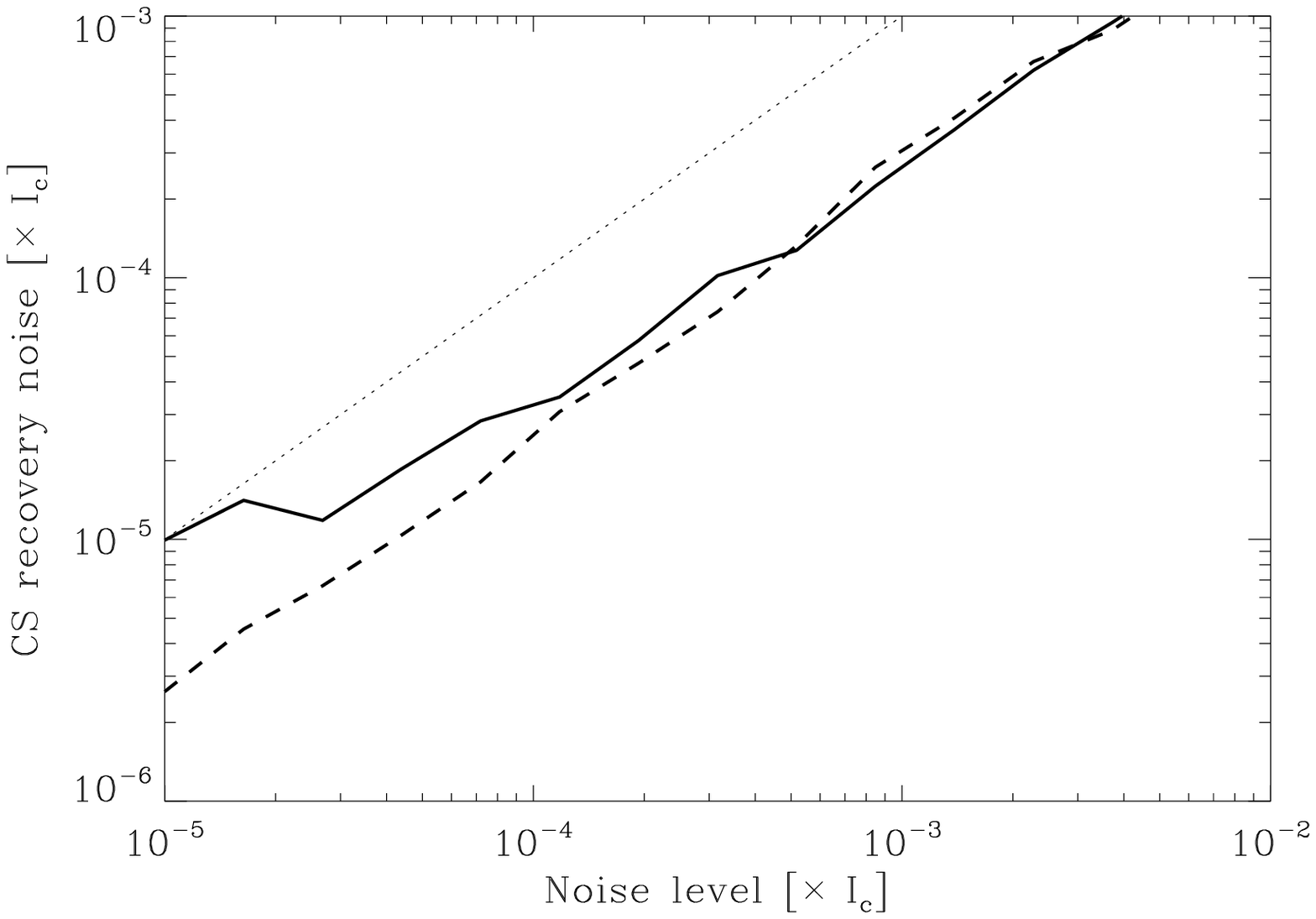}
\caption{Noise level of the CS reconstructed signal versus the noise level of each individual measurement.
The left panel shows the reconstruction with a wavelet basis set while the right panel has been obtained
using a universal PCA basis set. The sparsity in the wavelet case is assumed to be 12, while this number
is reduced to 5 for the PCA case. The dashed line shows the reconstruction error when only noise is
taken into account. The dotted line is the diagonal, the expected noise level if a standard spectrograph
is used. For comparison, the maximum amplitude of the signal is 4$\times$10$^{-4}$, so that
a noise level of 10$^{-4}$ gives a signal-to-noise ratio of 4.}
\label{fig:noise_reconstruction}
\end{figure*}

To test the influence of noise on our ability to recover the spectrum from the linear combinations
given by the sensing matrix, we analyze a synthetic case. We have chosen the spectral line at 6302 \AA\
for its widespread use. A very simplistic Stokes $I$ profile is built using a Voigt function tweaking the
width and depth to fit the average profile of the solar atlas \citep{wallace_atlasvis98}.
The Stokes $V$ profile emerging from a magnetized atmosphere is built under the weak-field approximation, in which
it is proportional to the wavelength derivative of Stokes $I$ \cite[e.g.,][]{landi_landolfi04}. We used
a magnetic flux density of 1 Mx/cm$^{-2}$, resulting in an amplitude
of 4$\times$10$^{-4}$ in units of the continuum intensity. The small value of the magnetic flux density is used intentionally so that the
observation is close to the detection limit of present spectro-polarimeters on relatively short exposure
times. We simulate observations with noise added and we use a Gaussian random sensing matrix with zero
mean and inverse variance equal to the assumed sparsity. We carry out experiments, shown in 
Fig. \ref{fig:noise_reconstruction}, using Daubechies-8 wavelets assuming that only 12 elements of the 
recovered vector are non-zero (left panel) and a universal PCA basis set assuming that the sparsity
of the vector is 5 (right panel). The PCA basis is obtained from a dataset observed with Hinode. 
The number of measurements is set to six times the sparsity level in each case. The 
solid line in each plot indicates the standard deviation of the difference between the exact profile and 
the reconstructed one for each value of the noise level. The dashed line is the same quantity but calculated
only for the noise. This curve is used to give an idea of the de-noising abilities of the decomposition on
a wavelet/PCA basis. When the signal-to-noise ratio (SNR) is very poor, in both experiments we see that a reduction of almost an
order of magnitude in the noise, clearly stating that the signal is well below the noise level. Only when the
SNR approaches $\sim$0.4, we see that the CS detects signal and gives a much better behavior than just the
direct measure of the profile using standard techniques. The reason has to be found on the fact that sparsity
is promoting the detection of signals contrary to the detection of noise. Noise is not sparse
in any of the used basis sets and the reconstruction is made assuming sparsity on the solution. This
is an important information that we are including as a prior on the CS recovery, something that is not
done in standard measurements.

\subsection{CS with polarimetric modulation}
Since existing instruments are not directly sensitive to polarization in the optical and infrared 
spectral domains, it is customary to use modulation schemes for measuring the Stokes parameters.
In some sense, modulation is another form of multiplexing which is carried out using 
retarders and polarizers. The monochromatic Stokes vector entering the telescope $\mathbf{S}$ is modulated
$N_\mathrm{mod}$ times for generating intensities that are linear combinations of the Stokes parameters of 
$\mathbf{S}$:
\begin{equation}
\mathbf{I}^\mathrm{out} = \mathbf{O}' \mathbf{S},
\end{equation}
where each row of the $N_\mathrm{mod} \times 4$ matrix $\mathbf{O}$ equals the first row of the Mueller matrix of each
modulation state. 

The obvious question is whether it is possible to apply CS for compressing the wavelength information
while still carrying out the modulation for detecting the four Stokes parameters. The answer is that
it is possible since both multiplexing operations "act" on different spaces: polarimetric modulation
applies to monochromatic Stokes parameters and the sensing matrix of CS applies to the wavelength
variation of a single Stokes parameter. Therefore, under the assumption that the polarimetric modulation 
is achromatic over the observed spectral range, their effect can be interchanged and
one ends up with solving four problems of the kind:
\begin{equation}
\mathbf{y}_i = \mathbf{\Phi} \mathbf{W}^T \mathbf{x}_i + \mathbf{e},
\label{eq:sensing_demodulated}
\end{equation}
where $\mathbf{x}_i$ is the demodulated wavelength variation of the $i$-th Stokes parameter 
\citep{deltoro_collados00} and we have assumed that the same CS matrix is used for all the
modulation states of the polarimeter. Consequently, the procedure to follow is to obtain the $N$ linear
measurements for each position of the polarimeter, thus leading to a set of $N N_\mathrm{mod}$ measurements. 
Then, following \cite{deltoro_collados00}, the pseudo-inverse of the modulation matrix $\mathbf{O}$ is 
applied to the $N_\mathrm{mod}$ modulation states of each linear measurements. At the end,
the four CS problems of Eq. (\ref{eq:sensing_demodulated}) are solved.

\subsection{Fringes and other spurious signals}
Among others, non-polarized and polarized fringes are undesirable contaminations present in many spectro-polarimetric
observations \citep[e.g.,][]{semel_fringe03}. Observations are partially cleaned from these fringes using
flat-fielding techniques. The remaining fringes are filtered out at the end of the reduction process 
with the disadvantage of having a large subjective 
component. Under the compressive sensing scheme these spurious signals are measured together with the real signal. It 
is a matter of the reconstruction to avoid introducing them into the final result, be it rejecting or
reconstructing them together with the true signal. Obviously, the ideal situation is to employ a fringe-free spectro-polarimeter 
to obtain a better reconstruction. We defer the deep investigation of this issue to a later study. However, we want 
to point out that preliminary experiments indicate that the reconstruction can efficiently reduce the amplitude of 
periodic fringes if the basis set used is not able to reproduce them. The drawback is that the ensuing reconstruction 
is less accurate than in the absence of fringes. Such a situation arises when applying a PCA basis set that is 
able to efficiently describe the spectro-polarimetric signals but not the periodic fringes. Another possibility 
of investigation is to assume that the fringes are sparse in the Fourier domain, carrying out the reconstruction 
merging the Fourier basis set and the basis set for the signal together.

\section{Applications}
\label{sec:other}

\subsection{Efficient Spectro-imagers}
Recent interest in Hadamard techniques for spectro-polarimetry \citep[see][for more details]{harwit_sloane79} 
can greatly benefit from CS techniques. 
Hadamard techniques allow to condensate spectral information inside single detector pixels through multiplexing. 
Before going into the use of these techniques in the framework of CS, it is advisable to describe these 
Hadamard techniques in spectro-polarimetry and the advantages they bring up through two illustrative examples. In  
traditional spectroscopy a certain amount of pixels (often a full dimension of a detector array) 
are dedicated to measure intensities at different wavelengths for the same point in an image. 
In Hadamard techniques, that is substituted by a temporal modulation over the Hadamard cyclic 
mask in a single detector pixel. Such exchange has two interesting applications:  in long-slit 
spectroscopy, the spectrum can be multiplexed in a single pixel and instead of using a 
2-dimensional detector array one can use a one dimensional array with increased 
acquisition cadences that allow for seeing freezing 
during the modulation cycle of polarimetry with the resulting improvement on polarimetric 
sensitivity \citep[see, e.g., ZIMPOL;][]{povel01}.

In double-pass substractive spectroscopy \citep{mein_msdp02} the resulting image has been filtered by a narrow spectral slit 
that selects a single wavelength per
pixel. Due to the dispersion of the first pass over the diffraction grating, the selected wavelength changes as one 
moves over the spatial image. To reconstruct the full 3D data cube with both spatial dimensions and spectral 
covering, every pixel has to be scanned over a range of wavelengths.  
Through the use of Hadamard cyclic matrices one can have several 
wavelengths sampled simultaneously in every pixel. As a result the temporal coherence of the recovered spectra is 
increased as several wavelengths over the spectral domain are detected simultaneously. Also, as a side effect, the 
raw images do not show any evident spectral features, what makes them 
more suitable for image reconstruction 
techniques

Such applications of Hadamard techniques to spectro-polarimetry are however hindered by 
the known fact that, in the presence of a multiplicative noise as photonic noise, the Hadamard 
transformation results in a reduction of SNRs with respect to the case of equivalent 
exposures without multiplexing \citep{harwit_sloane79}. The reduction in the SNR can be limited with 
the use of appropriate binary masks \citep{wuttig05}, although it is never too large for the usual cases 
in solar spectro-polarimetry.


Compressive sensing can help mitigate the problem. Since the Hadamard technique is applied to the 
spectral information, one can make use of the fact that there is prior information about the spectrum
to be measured and that, in consequence, the space of spectral profiles (with polarization included) 
is sparse, as demonstrated in the previous sections of this work. The use of the CS techniques 
illustrated above allows the recovery of the full spectral information with just a few spectral 
measurements. In the language of Hadamard techniques, this translates
into the fact that not all the data
acquisitions attached to the cyclic Hadamard masks are required for the recovery of the spectrum.
If traditionally a Hadamard mask of dimension $N$ would require $N$ cyclic measurements to solve the 
multiplexing linear system, CS techniques may be used to solve the system with just $MK$ acquisitions,
where $K\ll N$ is the sparsity of the signal (perhaps $K\sim0.1N$ at most) and $M>1$ is a small number. 
If each exposure lasts for $t_\mathrm{exp}$, the reduction in the required time for spectral information 
retrieval (from $N\times t_\mathrm{exp}$ to $MK \times t_\mathrm{exp}$) can then be used, not only to accelerate the 
full process of measurement, but also to repeat $N/(MK)$ times the same measurement and add them 
to gain a factor $\sqrt{N/(MK)}$ in signal to noise ratio. From the 
tests of  previous sections we can conclude that a figure of $K = 0.1 N$ is sufficient 
for the seeked precisions of the measurement. With a factor $M \sim 5$, the repetition of the measurements with a reduced 
Hadamard cycle can be used to gain roughly a factor $\sqrt{10/M} \sim 1.4$ in SNR. Such a factor would largely 
compensate the loss of SNR inherent to the use of the Hadamard techniques with a multiplicative 
photon noise. We conclude that the use of compressive sensing is strongly recommended for a 
successful application of Hadamard techniques to spectro-polarimetry.

\begin{figure}[!b]
\includegraphics[width=\columnwidth]{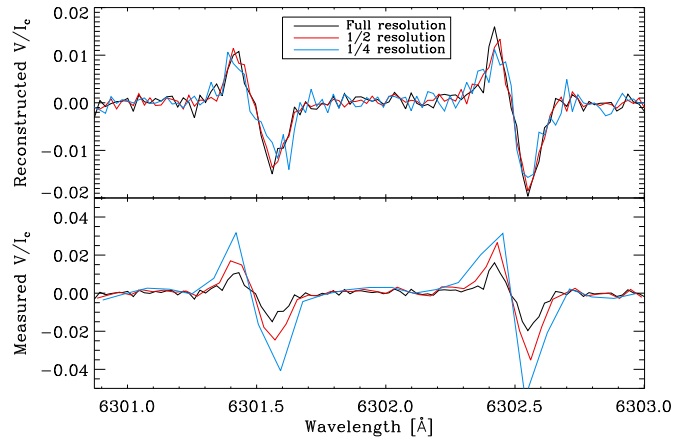}
\caption{Example showing how reliable reconstructions of high-resolution signals can be obtained from rebinned
data, provided that the signal is considered to be compressible. The upper panel shows the original (black)
and reconstructed profiles using $1/2$ (red) and $1/4$ (blue) of the original resolution. The data is
reconstructed with 5 PCA eigenvectors of a universal basis set. The lower panel shows the measurements
from which the reconstructions are obtained. The original signal is shown in black, while red and blue
lines show measurements rebinned to $1/2$ and $1/4$ of the original resolution.}
\label{fig:subnyquist}
\end{figure}

\subsection{Sub-Nyquist Spectrograph}
When prior information about the expected signals is available, the possibility of a spectrograph
sampling it is possible to think of a spectrograph able to measure the wavelength variation of the
Stokes parameters using resolution elements (pixels in the camera) larger than the spectral sampling.
In such a case, if one measures with a camera containing $n_\mathrm{pix}$ pixels, each one integrating $k$ spectral 
sampling steps, the sensing matrix of size $n_\mathrm{pix}\times kn_\mathrm{pix}$ can be written:
\begin{equation}
\Phi_{ij} = 
\left\{
\begin{array}{lr}
1  :& ki+1 < j < k(i+1)\\
0  :& \mathrm{otherwise.}
\end{array}
\right.
\end{equation}
This sensing matrix is probably not very efficient for reducing to the optimal value the number of
CS measurements but it suffices for our aims, since we are typically interested in cases where $k$ is 
not very large. An example of this is shown in Fig. \ref{fig:subnyquist}. A Stokes $V$ profile observed
with Hinode, shown in black lines in both panels is rebinned to $1/2$ and $1/4$ the original resolution by adding
two/four consecutive pixels together. The corresponding measurements are shown in the lower panel with
red and blue lines, respectively. Using five PCA eigenvectors of the universal set discussed in section 
\ref{sec:sparsity_zeeman}, the signals are reconstructed solving the $\ell_1$ optimization problem. The
reconstructed signals are shown in red and blue lines in the upper panel of Fig. \ref{fig:subnyquist},
corresponding to $1/2$ and $1/4$ of the original resolution, respectively. 

We point out that, if the signal is known to be sparse in the Fourier basis, one could consider that the
Nyquist-Shannon theorem should be applied to the frequency support where the signal is defined in the 
frequency domain. This is the case of signals for which the power associated with frequencies above a certain 
threshold are associated to noise. In such a case, one can use this new threshold, using the Nyquist-Shannon
sampling theorem, to estimate the number of pixels per resolution element. Reconstruction should be done
using the appropriate prior information. We have verified, although not shown here, that good 
reconstructions can be obtained using $1/2$ and $1/4$ of the original information by employing the Fourier
basis set and forcing the signal in the Fourier domain to be sparse. This has the advantage over PCA eigenvectors
that they are fully universal.

\begin{figure}
\includegraphics[width=\columnwidth]{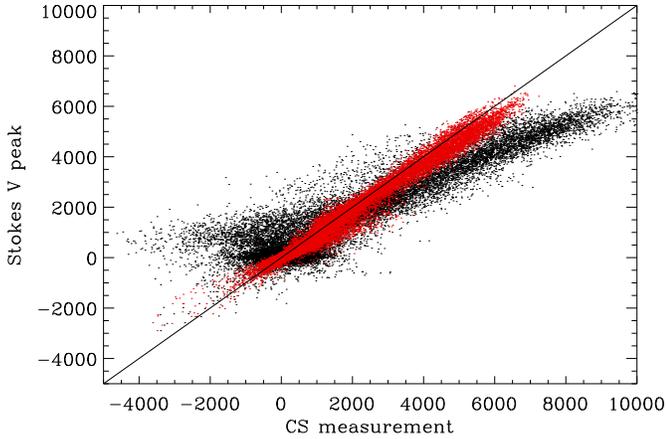}
\caption{Comparison between the measured Stokes $V$ amplitude and the one measured using CS techniques
with a Hadamard sensing matrix. The black dots show the comparison when only one measurement is done, while
the red dots present the results using 8 measurements.}
\label{fig:hadamard_pca_magnetometer}
\end{figure}

\subsection{Hadamard-PCA Magnetometer}
The combination of CS ideas and standard PCA techniques can be applied to develop
an efficient magnetometer. To this end, let us assume that the Stokes $V$ profile is well represented using
the first PCA eigenvector $v_1(\lambda)$ obtained empirically from a set of previous observations:
\begin{equation}
V(\lambda) \approx A v_1(\lambda).
\label{eq:decomposition}
\end{equation}
This equation assumes, therefore, that the observed Stokes profiles are 1-sparse in the PCA basis set, so 
that one can recover the signal, according to the CS theory, using of the order of 4-6 measurements. Using a 
sensing matrix $\mathbf{\Phi}$, we end up with the following measurement process:
\begin{equation}
y_i = \sum_j \Phi_{ji} V(\lambda_j), \qquad \forall i=1,\ldots,N_\mathrm{meas}
\end{equation}
If the sparsity constraint is used, the proportionality constant $A$ can be obtained from the observations with
a linear fit:
\begin{equation}
A \approx \frac{\sum_i y_i f_i}{\sum_i f_i^2},
\end{equation}
where $f_i=\sum_j \Phi_{ji} v_1(\lambda_j)$
Note that a similar result could have been obtained if we measure projections of the data over the
PCA eigenvectors. However, this presents the difficulty of its practical implementation because the PCA
eigenvectors contain negative values. Measuring with sensing matrices like a binary (1/0) Hadamard matrix 
that is incoherent with the PCA eigenvectors leads to a universal sensing process. We point out that
only one measurement gives a rough estimation of the magnetometer signal although it should be used with care.
An example of the capabilities of the method is shown in Fig. \ref{fig:hadamard_pca_magnetometer}. A scan of 
a sunspot obtained on 27 February 2007 with the SOT/SP onboard Hinode has been used. The vertical axis
presents the value of the Stokes $V$ profile at a fixed wavelength corresponding to the blue
$\sigma$ component of the 6302.5 \AA\ line. The horizontal axis presents the value inferred from the CS
measurements. The black dots show the scatter when only one measurement is done, while the red dots
show what happens when 8 measurements are done.

We point out that, if enough measurements are taken, it is possible to include more eigenvectors
in the decomposition of Eq. (\ref{eq:decomposition}) and obtain information about the projection
along them from the observations. If Stokes $I$ is decomposed, continuum images and velocities 
can be inferred from the first and second eigenvectors, respectively. Likewise, if Stokes $V$ is used,
magnetic flux and magnetic velocities can be inferred.

\section{Conclusions}
\label{sec:conclusions}
This paper demonstrates the feasibility of applying compressive sensing techniques for measuring
the wavelength variation of the Stokes parameters observed in stellar atmospheres. We have shown
that spectro-polarimetric signals are, in general, compressible on universal basis sets. However,
in general, it is more advantageous to use empirical basis sets like that obtained from 
principal component analysis because data can be more efficiently reproduced on such a 
basis. We stress that the results presented in this paper are extensible to standard spectroscopic observations
and not only to linear and circular polarization profiles, so that any day-time and night-time spectrograph
can take advantage of these techniques.

According to our results, it is possible to measure Stokes parameters using less than a half
of the measurements one should carry out when strictly applying the Nyquist-Shannon sampling theorem.
Compressing sensing leads to several interesting effects. Since less measurements are made,
an inherent reduction in the exposure time is present. Such reduction can be used to do more
measurements in the same total time, thus allowing less noisy observations. The a-priori information
encoded in the sparsity condition results in the fact that the reconstructed signal is much less noisy than 
one should expect. The reason is that filtering is applied simultaneously while measuring. For instance,
several of the examples shown in this paper produce signals that are automatically filtered with the principal component analysis 
applied by \cite{marian08,marian_pcafilter08}.

We have proposed potential applications of CS to the field of spectro-polarimetry, testing them
numerically on real data. Some of these techniques can be straightforwardly applied to existing
instruments, while other proposals need more profound modifications. The future of observational 
spectro-polarimetry, at least in solar physics, has to be rooted on the development of 
two-dimensional spectro-polarimeters. Since detectors are only two-dimensional at the moment,
scanning schemes have to be used. We consider that double-pass substractive spectroscopy
constitutes a very appealing technique for two-dimensional spectro-polarimetry if combined with 
multiplexing techniques using Hadamard masks. Compressive sensing will help reduce significantly the 
total exposure time, thus allowing an increase in the final SNR for a fixed integration time.

Because of the natural physical interpretation of projections of the observed Stokes profiles along PCA 
eigenvectors \citep{skumanich02}, it would be desirable to directly measure such projections.
Using compressive sensing techniques, we have proposed a technique that, thanks to Hadamard masks, is able 
to retrieve such projections from an universal multiplexing. The advantage is that this multiplexing is 
binary and easy to build.

Finally, we analyze the plausibility of a spectro-polarimeter that does not fulfill the Nyquist-Shannon
sampling theorem. Some a-priori information about the expected signals is available and compressive
sensing techniques can take full advantage of this information for recovering the signals
from a reduced set of measurements. We show with an example that it is possible to reconstruct
Stokes profiles using the information obtained from adding the signal in consecutive pixels. In some
sense, this can be understood as a super-resolution scheme in which one knows the basis set in which the
high-resolution signal can be efficiently developed.

\begin{acknowledgements}
We thank Rafael Manso Sainz and Mar\'{\i}a Jes\'us Mart\'{\i}nez Gonz\'alez for illuminating discussions and
carefully reading the manuscript.
Financial support by the Spanish Ministry of Education and Science through project AYA2007-63881 is gratefully acknowledged.
\end{acknowledgements}

\begin{appendix}
\section{Compressive sensing}
\label{sec:appendix}
As noted in the main text, the multiplexing scheme for a sparse signal reads:
\begin{equation}
\mathbf{y} = \mathbf{\Phi} \mathbf{W}^T \mathbf{x} + \mathbf{e},
\end{equation}
with the condition that $\mathbf{x}$ is sparse. When the number of measurements is
much smaller than the size of the signal, the sparsest
solution fulfills:
\begin{equation}
\min_{\mathbf{x}} \parallel \mathbf{x} \parallel_0 \textrm{subject to} 
\parallel \mathbf{y} - \mathbf{\Phi} \mathbf{W}^T \mathbf{x} \parallel_2 < \epsilon,
\label{eq:l0_minimization_APP}
\end{equation}
where $\parallel \mathbf{x} \parallel_0$ is the $\ell_0$ pseudo-norm that equals the number of non-zero elements of the vector
$\mathbf{x}$. Since solving this problem is, in general, not feasible, it has been
demonstrated by \cite{candes06,candes_2_06} that, if the matrix $\mathbf{\Phi} \mathbf{W}^T$ fulfills 
the Restricted Isometry Property \citep[RIP;][]{candes_2_06}, the solution to the problem
\begin{equation}
\min_{\mathbf{x}} \parallel \mathbf{x} \parallel_1 \textrm{subject to} 
\parallel \mathbf{y} - \mathbf{\Phi} \mathbf{W}^T \mathbf{x} \parallel_2 < \epsilon,
\label{eq:l1_minimization_APP}
\end{equation}
is equivalent to that of Eq. (\ref{eq:l0_minimization_APP}). We note in passing that
the RIP condition is a sufficient condition and it is often too restrictive. The advantage of the
last problem is that it can be easily solved using linear programming techniques.

Intuitively, the RIP condition states that the action of the 
$\mathbf{\Phi} \mathbf{W}^T$ operator on the sparse vector $\mathbf{x}$ does not
modify excessively its $\ell_2$ norm. Mathematically:
\begin{equation}
(1-\delta_K) \parallel \mathbf{x} \parallel_2^2 ~\leq~ \parallel \mathbf{\Phi} \mathbf{W}^T \mathbf{x} \parallel_2^2
~\leq~ (1+\delta_K) \parallel \mathbf{x} \parallel_2^2
\end{equation}
for all $K$-sparse vectors $\mathbf{x}$ and $\delta_K < 1$. 

In spite of the mathematical importance of the RIP condition, it is more intuitive to
think on terms of coherence between the sensing matrix and the transformation matrix.
In general, for a sensing matrix to be considered good, it should be as incoherent as possible
with the transformation matrix. Every row of the sensing matrix should be able to obtain as much
information as possible from the sparse vector $\mathbf{x}$ in order to facilitate its
reconstruction with as few measurements as possible. This is achieved when the sensing
matrix and the transformation matrix are as incoherent as possible. The coherence
between the two matrices is defined as \citep{candes07}:
\begin{equation}
\mu(\mathbf{\Phi},\mathbf{W}) = \max_{\phi \in \mathbf{\Phi}, w \in \mathbf{W}} | \langle \phi, w \rangle |,
\end{equation}
where $\phi$ and $w$ are, respectively, columns and rows of the matrices $\mathbf{\Phi}$ and $\mathbf{W}$.
Quite generally \citep{candes07}, for a sensing matrix of size $N\times M$, it is possible
to recover an $s$-sparse vector using a number of linear combinations that fulfills:
\begin{equation}
M \geq C \mu(\mathbf{\Phi},\mathbf{W})^2 K \log N,
\label{eq:nmeas}
\end{equation}
where $C$ is a constant of order 1. According to this equation, if one is able to find
sensing matrices with small coherence with respect to the basis set of interest, one should
be able to recover the sparse signal using a number of measurements that is proportional
to $K \log N$. As we have shown in the main text, the proportionality constant is typically between 4 and 6.

\subsection{Recovery algorithm}
\label{sec:recovery_IHT}
For the recovery problem in our experiments, different methods have
been developed during the last few years. After testing several methods, we found that the recent algorithm
presented by \citep{iht_blumensath08} is very efficient in terms of computing time and shows state
of the art performance. The method
uses the simple iterative procedure:
\begin{equation}
\mathbf{x}^{n+1} = H_s \left[ \mathbf{x}^{n} + \mu \mathbf{W} \mathbf{\Phi}^T \left( \mathbf{y} - 
\mathbf{\Phi} \mathbf{W}^T \mathbf{x}^n \right) \right],
\label{eq:IHT}
\end{equation}
where $H_s(\mathbf{t})$ is a non-linear thresholding operator that leaves as non-zero the $s$ elements of the
vector $\mathbf{t}$ with the largest absolute value, setting to zero the rest of elements. The method
is guaranteed to be stable thanks to the re-scaling quantity $\mu$. Indeed, according
to our experience, its stability is remarkable, converging to the
solution in almost all experiments. The method is 
initialized by $\mathbf{x}^0=0$. The main drawback of the method (usually common to all recovery methods) 
is that the sparsity of the solution, $s$, has to be chosen in advance. The advantage is that only multiplications with the
matrices $\mathbf{\Phi}$ and $\mathbf{W}$ (and their transposes) are needed. Many sparsity-promoting basis sets
are accompanied by fast multiplication algorithms (e.g., fast fourier transform, fast wavelet transform, etc.).
In such a case, the computing time of the multiplication with the $\mathbf{W}$ and $\mathbf{W}^T$ matrices scales
as $O(n)$ for the fast wavelet transform and as $O(n \log n)$ for the fast fourier transform, instead of scaling
as $O(n^2)$ like in a standard matrix-vector product.

\end{appendix}


\end{document}